\documentclass[epj, final, nopacs]{svjour}

\def\makeheadbox{\relax}

\usepackage{graphicx}
\usepackage{amsmath}
\usepackage{upgreek}
\usepackage{amssymb}
\usepackage{xcolor}
\usepackage{microtype}
\usepackage{hyperref}
\usepackage{newtxtext, newtxmath}
\usepackage{xurl} 
\hypersetup{colorlinks=true, allcolors=blue}
\usepackage[switch]{lineno} 
\usepackage{orcidlink}

\usepackage{filecontents}

\newcommand{\ns}{n_\text{s}}
\newcommand{\pknmu}{P(k \mid N_{\upmu})}
\newcommand{\pkmu}{P(k \mid \mu)}
\newcommand{\nmu}{N_{\upmu}}
\newcommand{\muhat}{\hat{\mu}}
\newcommand{\mutilde}{\tilde{\mu}}
\newcommand{\secref}[1]{Sec.~\ref{#1}}
\newcommand{\figref}[1]{Fig.~\ref{#1}}
\newcommand{\eqnref}[1]{Eq.~(\ref{#1})}

\newcommand{\appref}[1]{Appendix~\ref{#1}}

\begin{document}

\title{A generalized likelihood model for segmented muon counters}

\author{
    J.~de~Jes\'us\orcidlink{0000-0002-4741-1769}\inst{1,2, 3,}\thanks{e-mail: joaquin.dejesus@usc.es (corresponding author)}  \and
    J.~M.~Figueira\orcidlink{0000-0002-6768-5214}\inst{1} \and
    F.~S\'anchez\orcidlink{0000-0002-6861-6261}\inst{1} \and
    D.~Veberič\orcidlink{0000-0003-2683-1526}\inst{2}
}

\institute{
    Instituto de Tecnolog\'ia en Detecci\'on y Astropart\'iculas (CNEA, CONICET, UNSAM), Buenos Aires, Argentina
    \and 
    Karlsruhe Institute of Technology (KIT), Institute for Astroparticle Physics, Karlsruhe, Germany 
    \and 
    Instituto Galego de F\'isica de Altas Enerx\'ias (IGFAE),
    R\'ua de Xoaqu\'in D\'iaz de R\'abago, s/n, Campus Vida,
    Universidade de Santiago de Compostela, 15705, Santiago de Compostela, Galicia, Spain
}

\date{August 2026} 

\abstract{
Measurements of the muonic component of extensive air showers constrain cosmic-ray mass composition and hadronic interactions at energies beyond those accessible at accelerators. 
Arrays of segmented detectors with binary readout are widely used for this purpose: they sample the muon density at different distances from the shower core to reconstruct the muon lateral distribution function (LDF). 
Each detector response is summarized by the number of activated segments, $k$, whose probability distribution provides the likelihood relating the observation to the expected muon content. 
Signal pile-up, detector inefficiency, corner-clipping muons, and background signals shape this distribution, and neglecting them can bias the reconstruction.
Existing analytical models include pile-up but otherwise assume an ideal detector response.
We develop a unified statistical framework that incorporates inefficiency, corner clipping, and background through a small set of physically interpretable parameters. 
We derive exact expressions for the detector response and the likelihood required for muon-LDF reconstruction, together with a simple binomial approximation that preserves the main statistical properties of the exact distribution. 
Dedicated Monte Carlo simulations are used to assess the impact of the assumptions underlying the analytical treatment and show that it is negligible over the parameter range considered. 
They also show that the exact and approximate likelihoods yield similar performance in terms of estimator bias and confidence-interval coverage. 
Although motivated by the Underground Muon Detector of the Pierre Auger Observatory, the framework applies more broadly to segmented particle detectors with binary readout in which particle content is inferred from the number of activated segments.
}

\maketitle


\section{Introduction}
\label{sec:intro}

Ultra-high-energy cosmic rays ($E > 10^{18}$ eV) are the most energetic particles known in the Universe, reaching energies far beyond those accessible with human-made accelerators.
Yet, despite decades of observations, their sources and the mechanisms governing their acceleration and propagation remain open questions.
The evolution of their mass composition with energy provides essential constraints on possible source populations and helps discriminate among competing acceleration and propagation scenarios.

At these energies, the mass composition cannot be measured directly and is instead inferred from the extensive air showers produced by cosmic rays in the atmosphere.
Two key mass-sensitive observables are the atmospheric depth of shower maximum, $X_\text{max}$, and the muon content at ground level.
The interpretation of these air-shower observables in terms of mass composition  relies on comparisons with simulations based on hadronic interaction models constrained by accelerator data but extrapolated to energies and kinematic regions beyond those directly accessible at colliders.

In this context, precise muon measurements serve two closely connected purposes.
First, current hadronic interaction models do not provide a consistent description of $X_\text{max}$ and muon measurements, indicating that the modeling of hadronic interactions at ultra-high energies remains incomplete.
For primary compositions compatible with $X_\text{max}$ observations, simulations generally predict fewer muons than are observed in air showers, a tension commonly referred to as the \emph{muon puzzle}~\cite{ArteagaVelazquez2023WHISP,Albrecht2022MuonPuzzle}.
Resolving this discrepancy requires precise muon data, which provide key experimental constraints for ongoing efforts to develop and tune next-generation hadronic interaction models using combined accelerator and air-shower data~\cite{Albrecht2026GlobalTuning}.
Improved models would, in turn, allow the cosmic-ray mass composition to be inferred more reliably from muon observables.

Second, ground-based muon detectors operate with a much higher duty cycle than the fluorescence telescopes used to measure $X_\text{max}$ and can therefore provide substantially larger event samples.
Once the muonic component can be modeled reliably, these measurements will enable high-statistics studies of the cosmic-ray mass composition.
Precise measurements of air-shower muons are thus central both to improving our understanding of hadronic interactions at ultra-high energies and to exploiting the statistical power of ground-based detectors for composition studies.

A common approach to measuring the muonic component of extensive air showers is to use shielded segmented detectors, in which the sensitive area is divided into segments that are read out independently~\cite{Borione1994,Chiba1992,Hayashida1995,Aab2016AMIGAPrototype}.
The shielding strongly suppresses the more abundant electromagnetic component of the shower, allowing the muonic component to be measured separately.
In detectors with binary readout, in which each segment is classified as either activated or non-activated, the basic observable is the number of activated segments, denoted by $k$, where an activated segment is one that records a signal satisfying a detector-specific activation criterion, such as exceeding a predefined threshold.

The relation between $k$ and the number of impinging muons, $\nmu$, is inherently probabilistic because several mechanisms, hereafter collectively referred to as \emph{detector effects}, can cause $k$ to differ from $\nmu$.
These include \emph{pile-up}, where multiple muons hit the same segment; \emph{corner clipping}, where a single muon activates two adjacent segments; \emph{detector inefficiency}, where a muon fails to activate any segment; and \emph{background signals}, which can activate segments even in the absence of air-shower muons.
The detector response is therefore naturally described by $\pknmu$, the probability of observing $k$ activated segments when exactly $\nmu$ muons impinge on the detector.

In practice, however, the reconstruction of the air-shower muon content does not aim to determine an independent value of $\nmu$ at each detector.
An array of muon detectors samples the shower at different distances from the core, and the measurements are combined to estimate the parameters of the muon lateral distribution function (LDF), $\rho(r;\mathbf{p})$, which describes the muon density as a function of the distance $r$ from the shower core, measured in the shower plane.
For a horizontal detector of area $A$, with the muon density defined in the shower plane, the LDF predicts an expected number of impinging muons
\begin{equation}
\label{eq:mu_expected}
\mu = \rho(r;\mathbf{p}) A \cos\theta,
\end{equation}
where $\theta$ is the shower zenith angle and $\mathbf{p}$ denotes the LDF parameters.
The actual number of impinging muons, $\nmu$, is then modeled as a Poissonian random variate with mean $\mu$ and probability mass function $P(\nmu \mid \mu)$.

The distinction between $\nmu$ and $\mu$ can be illustrated by considering two identical detectors located at the same distance $r$ from the shower core in the same air-shower event.
The LDF assigns them the same expected number of impinging muons, $\mu$, but they will generally be crossed by different numbers of muons because of Poisson sampling fluctuations.
Since the LDF predicts $\mu$, rather than a particular realization $\nmu$, its likelihood-based reconstruction requires the distribution $\pkmu$.

This distribution must account both for the detector response at fixed $\nmu$ and for fluctuations in the actual number of impinging muons.
It is obtained by marginalizing $\pknmu$ over all possible values of $\nmu$,
\begin{equation}
\label{eq:prob_mu_marginalization}
\pkmu
=
\sum_{N_\mu=0}^{\infty}
\pknmu 
\,P(N_\mu \mid \mu).
\end{equation}
For an array of detectors, each measurement contributes a likelihood term $\pkmu$, evaluated at the value of $\mu$ predicted by the LDF for the detector position.
The distribution $\pkmu$ therefore provides the fundamental likelihood ingredient for reconstructing the muon LDF with an array of segmented detectors.

Expressions for $\pknmu$ and $\pkmu$ for ideal segmented detectors were derived in Refs.~\cite{supa2021} and~\cite{ravignani2015}, respectively, accounting only for signal pile-up and assuming a fully efficient detector with no background or corner-clipping muons.
The central goal of this work is to extend these results by deriving a physically motivated analytical expression for $\pknmu$ that incorporates detector inefficiency, corner clipping, and background through simple probabilistic models with physically interpretable parameters, and to use it to obtain $\pkmu$ through the marginalization in \eqnref{eq:prob_mu_marginalization}.
While motivated by the Underground Muon Detector (UMD) of the Pierre Auger Observatory~\cite{Aab2016AMIGAPrototype}, the resulting framework applies more generally to segmented detectors in which the particle content is inferred from the number of activated segments.

This paper is organized as follows.
In \secref{sec:dist_nmu}, we introduce the detector-response model, derive an exact expression for $\pknmu$, and examine how the different detector effects modify its shape.
We also discuss the experimental interpretation of the model parameters, the assumptions underlying the framework, and possible extensions.
In \secref{sec:dist_mu}, we analytically marginalize over the Poisson fluctuations in $\nmu$ to obtain an exact expression for $\pkmu$, discuss its use as a likelihood for muon reconstruction, and examine how detector effects influence the dynamic range.
We then introduce a numerically stable and statistically transparent binomial approximation that provides a closed-form estimator for $\mu$.
In \secref{sec:monte_carlo_validation}, we use Monte Carlo simulations to assess the approximations underlying the analytical detector-response model and compare the exact and approximate likelihoods in terms of estimator bias and confidence-interval coverage.
Our conclusions are presented in \secref{sec:conclusions}.

\section{Probability distribution of $k$ for fixed $N_{\upmu}$}
\label{sec:dist_nmu}

In this section, we derive a probabilistic model for the number of activated segments, $k$, given a fixed number of impinging muons, $\nmu$.
We consider a detector divided into $\ns$ contiguous segments arranged in a one-dimensional sequence, such that each inner segment has two neighboring segments, whereas each edge segment has only one.
Each segment is read out independently. 
When an air-shower event triggers the detector, the response of each segment is recorded over a fixed readout window of duration $T$. 
Signals satisfying the detector-specific activation criterion are identified within this window, and a segment is classified as activated if at least one such signal is recorded. 
Multiple signals recorded in the same segment within the readout window therefore contribute only one activated segment.


We first consider an ideal segmented detector and then progressively incorporate the detector effects.
We examine how these effects modify the resulting probability distribution and finally discuss the main assumptions of the model, their implications, and possible extensions of the framework.

\subsection{Ideal segmented detector}

We begin by considering an \emph{ideal} segmented detector in which only the pile-up effect is taken into account.
In this idealization, corner-clipping muons are neglected, and a 100$\%$ efficient, background-free detector is assumed.
This provides a baseline model that will serve as the starting point for incorporating additional detector effects.

For illustrative purposes, we adopt the analogy introduced in Ref.~\cite{gesualdi2022}, in which the problem of counting particles in a segmented detector is mapped onto the classical occupancy problem, or “balls-in-boxes” experiment.
In this analogy, each ball represents a muon, with the total number of balls given by $\nmu$, while the boxes represent the $\ns$ detector segments.

A realization of the experiment consists of randomly assigning each ball (muon) to one of the $\ns$ boxes (segments). 
After the assignment, each box is either occupied (containing one or more balls, corresponding to a segment with at least one muon signal) or empty (containing no balls, corresponding to a segment with no muon signal).
The total number of occupied boxes corresponds to the number of activated segments, $k$.

Under these idealized conditions, $\pknmu$ is given by~\cite{supa2021}
\begin{equation}
\label{eq:prob_nmu_base}
P_\text{ideal}(k \mid \nmu) = {\ns \choose k} S(\nmu, k) \frac{k!}{n_{s}^{\nmu}},
\end{equation}
where $S(\nmu, k)$ is the Stirling number of the second kind, defined as
$S(\nmu, k) = \frac{1}{k!} \sum_{j=0}^k {k \choose j} (-1)^j (k-j)^{\nmu}$.

\begin{figure}
    \centering
    \def\w{0.48}
\includegraphics[width=\w\textwidth]{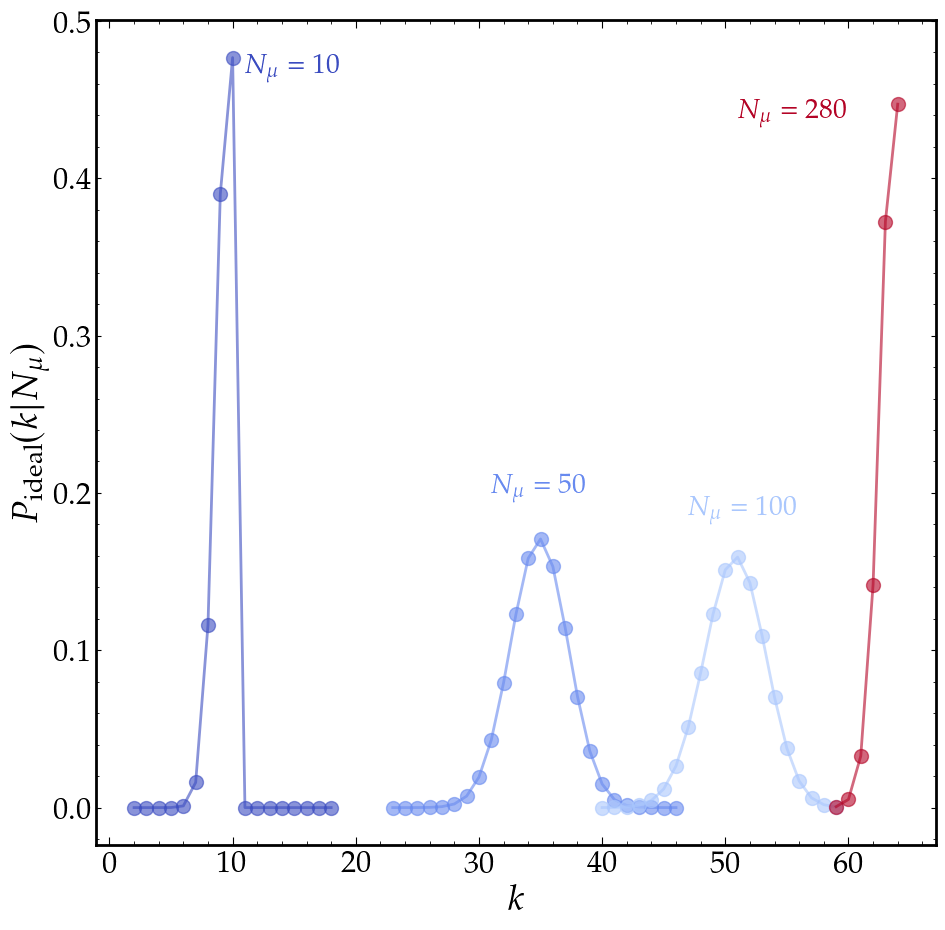}
    \caption{Probability distribution of $k$ for several values of $\nmu$ in an ideal segmented detector. A detector with $\ns=64$ segments, as the Underground Muon Detector of the Pierre Auger Observatory, is assumed.}
\label{fig:p_nmu_ideal}
\end{figure}

In \figref{fig:p_nmu_ideal}, $P_\text{ideal}(k \mid \nmu)$ is shown for several values of $\nmu$.
All plots in this paper assume a segmentation of $\ns = 64$, corresponding to the Underground Muon Detector~\cite{Aab2016AMIGAPrototype,PierreAuger:2021calibration,PierreAuger:2020MuonContent} of the Pierre Auger Observatory~\cite{PierreAuger:2015Observatory,PierreAuger:2020ObservatoryUpgrade}.
Since the number of activated segments cannot exceed the number of incident muons, 
$k$ is constrained by $k \leq \nmu$.
This effect is most evident when $\nmu$ is small compared to the number of segments 
$\ns$ (e.g., $\nmu = 10$).
In this regime, the distribution of $k$ is typically peaked near $\nmu$, indicating that each muon is likely to hit a different segment and that pile-up effects are negligible.
In contrast, for larger values of $\nmu$ (e.g., $\nmu= 50$ or $100$), pile-up becomes significant, and the distribution of $k$ peaks at values noticeably smaller than $\nmu$.
Finally, when $\nmu$ is much larger than the number of segments (e.g., $\nmu = 280$), the distribution saturates and peaks around $\ns$.

%

\subsection{Realistic segmented detector}
\label{sec:dist_nmu_real}

Real segmented detectors are affected by more than pile-up alone.
Detector inefficiency (muons failing to activate any segment), corner-clipping muons (muons activating two neighboring segments), and background signals (false positives, i.e., segments activated by atmospheric muons or by detector noise) all contribute to the distribution of $k$ and, if not properly accounted for, can bias the estimation of $\nmu$.

The effects of these mechanisms on $k$ do not simply add, because all resulting signals are subject to pile-up.
For instance, a corner-clipping muon produces two signals that can subsequently pile up with signals from other muons or from background.
Therefore, a consistent description of the detector response requires modeling all these effects jointly.

A key observation is that pile-up occurs at the level of signals rather than muons.
In the ideal detector, each of the $\nmu$ muons produces exactly one signal, so the number of signals, hereafter denoted by $M$, is fixed to $\nmu$.
In a realistic detector, however, $M$ is a stochastic quantity, determined by how many muons produce zero, one, or two signals, together with the contribution from background.

To account for this, we reformulate the problem in terms of $M$ rather than $\nmu$.
Inefficient muons reduce the number of signals available to pile up, while corner-clipping muons and background increase it.
In the balls-in-boxes analogy, inefficient muons remove balls from the system, whereas corner-clipping muons and background signals add extra balls.
Accordingly, we define

\begin{equation}
\label{eq:Neff}
M = \nmu - n_0 + n_\text{cc} + m,
\end{equation}
where $n_0$ and $n_\text{cc}$ denote the number of undetected and corner-clipping muons, respectively, and $m$ is the number of background signals.

For a given $(n_0, n_\text{cc}, m)$, there are $M$ signals available to pile up.
These $M$ signals are assumed to be uniformly distributed among the $\ns$ segments, following the same occupancy model as in \eqnref{eq:prob_nmu_base}, with $\nmu$ replaced by $M$.
This leads to the expression for the full distribution
\begin{equation}
\label{eq:prob_nmu_intermediate}
\pknmu = \!\!\! \sum_{(n_0, n_\text{cc}, m)} \!\!\! P(n_0, n_\text{cc}, m \mid \nmu) \, P_\text{ideal}(k \mid M),
\end{equation}
where $P(n_0, n_\text{cc}, m \mid \nmu)$ is the probability of the configuration $(n_0, n_\text{cc}, m)$, and the sum runs over all possible configurations of these variables for a given $\nmu$.

To model $P(n_0, n_\text{cc}, m \mid \nmu)$, we note that $n_0$ and $n_\text{cc}$ depend on the number of impinging muons $\nmu$, whereas $m$ is independent of $\nmu$.
Accordingly, the background contribution can be modeled separately.


When a single muon traverses the detector, three outcomes are possible: it produces signals in zero, one, or two (neighboring) segments. 
Let us denote by $p_0$ the probability that no signal is produced—i.e., the inefficiency probability—and by $p_\text{cc}$ the corner-clipping probability, corresponding to the production of signals in two neighboring segments. 
The probability of producing a signal in a single segment is then $p_1 = 1 - p_0 - p_\text{cc}$.
For $\nmu$ muons, the numbers of undetected muons $n_0$ and corner-clipping muons $n_\text{cc}$ follow a multinomial distribution with parameters $\nmu$ and probabilities $(p_0, p_\text{cc}, p_1)$ as
\begin{equation}
\begin{split}
\label{eq:multinomial}
P(n_0, n_\text{cc} \mid\nmu) &=
\frac{\nmu!}{n_0!\, n_\text{cc}!\, (\nmu-n_0-n_\text{cc})!} \times \\
&\quad \times
p_0^{n_0}\, p_\text{cc}^{n_\text{cc}}\,
\left(1 - p_0 - p_\text{cc}\right)^{\nmu-n_0-n_\text{cc}} .
\end{split}
\end{equation}

The probabilities $p_0$, $p_1$, and $p_\text{cc}$ generally depend on the muon trajectory, deposited energy, and impact point along the segment. 
As a result, different muons are, in principle, characterized by different values of these probabilities.
Throughout this work, $p_0$, $p_1$, and $p_\text{cc}$ should be interpreted as effective quantities, averaged over the distributions of these underlying variables.

To model the background signals, we denote by $R$ the rate of background signals in a single segment.
Assuming that background signals occur according to a Poisson process in time, the number of background signals in one segment during the readout window follows a Poisson distribution with mean $\lambda_1 = R \, T$.
For a detector with $\ns$ independent segments, the total number of background signals $m$ follows a Poisson distribution with mean $\Lambda = \ns \lambda_1$,
\begin{equation}
\label{eq:prob_noise}
P_\text{bg}(m) = \frac{\Lambda^m}{m!} e^{-\Lambda}.
\end{equation}
Assuming independence between the muon-induced and background contributions, we can write
\begin{equation}
P(n_0, n_\text{cc}, m \mid \nmu) = P(n_0, n_\text{cc} \mid \nmu)\, P_\text{bg}(m).
\end{equation}
Combining these ingredients, the full distribution can be written as
%
%
%
\begin{equation}
\begin{split}
\label{eq:prob_nmu_brute}
\pknmu ={}&
\sum_{m=0}^{\infty} \,
\sum_{n_0+n_\text{cc}\leq\nmu}
P_\text{bg}(m)\,
P(n_0,n_\text{cc}\mid\nmu)
\\
&\times
P_\text{ideal}
\left(k\mid\nmu-n_0+n_\text{cc}+m\right).
\end{split}
\end{equation}
where the inner sum runs over all $(n_0, n_\text{cc})$ pairs satisfying $n_0 + n_\text{cc} \leq\nmu$.
The sums over $m$ and  $(n_0, n_\text{cc})$  can be evaluated in closed form (see \appref{app:pknmu_simplified}), yielding
\begin{equation}
\begin{split}
\label{eq:prob_nmu}
\pknmu ={}&
{\ns \choose k} \sum_{j=0}^{k} (-1)^{k-j} {k \choose j} e^{-\ns\lambda_1(1-j/\ns)} \,\, \times  \\ 
&\times \left(p_0 + p_1 \frac{j}{\ns} + p_\text{cc}\left(\frac{j}{\ns}\right)^2 \right)^{\nmu} \!\!\!. 
\end{split}
\end{equation}
When setting $p_0 = p_\text{cc} = 0$ and $\lambda_1 = 0$, \eqnref{eq:prob_nmu_base} is recovered.

Within the assumptions stated in \secref{subsec:model_assumptions}, \eqnref{eq:prob_nmu} defines a complete statistical model for the response of a realistic segmented detector in terms of a small set of physically interpretable parameters $(p_0, p_\text{cc}, \lambda_1)$.
This set fully determines the distribution $\pknmu$, and therefore the statistical mapping between the true muon number $\nmu$ and the observed segment multiplicity $k$.
For applications in which the actual number of impinging particles $\nmu$ is itself the quantity of interest, \eqnref{eq:prob_nmu} can be used directly as a likelihood for $\nmu$, providing the basis for maximum-likelihood estimation and the construction of confidence intervals.
Practical use of this likelihood, however, requires the detector-response parameters $(p_0,p_\text{cc},\lambda_1)$ to be experimentally constrained. 
We therefore briefly discuss how these parameters can be determined in real detectors.

\subsection{Experimental interpretation and determination of the model parameters}

The model parameters can be constrained through dedicated detector measurements or directly from air-shower data.
In the laboratory, a muon telescope can be used to trigger the detector, allowing $p_0$ and $p_\text{cc}$ to be estimated from the fractions of events in which zero, one, or two neighboring segments are activated, with $p_1=1-p_0-p_\text{cc}$.
If track reconstruction is available, their dependence on the muon trajectory can also be characterized.
In air-shower data, the corner-clipping probability can be estimated from the characteristic timing signature of two quasi-simultaneous signals in neighboring segments~\cite{DeJesus2026}.
Since shower muons are strongly collimated around the shower axis, $p_\text{cc}$ can then be parameterized as a function of the shower direction, which provides an approximation to the muon direction.
The background parameter $\lambda_1$ can be determined either in the laboratory, by measuring the signal rate $R$ in a segment in the absence of a shower trigger, or directly from air-shower data using signals recorded outside the shower time window.


\subsection{Qualitative behavior of the model}

To gain insight into \eqnref{eq:prob_nmu}, we first consider a background-free detector with $\lambda_1 = 0$, in order to isolate the impact of detector inefficiency ($p_0$) and corner-clipping ($p_\text{cc}$) on the distribution.
Since the background contribution is independent of $p_0$ and $p_\text{cc}$, all qualitative features discussed in this subsection remain valid in the general case in which $\lambda_1 \neq 0$.

\begin{figure}
    \centering
    \def\w{0.48}
\includegraphics[width=\w\textwidth]{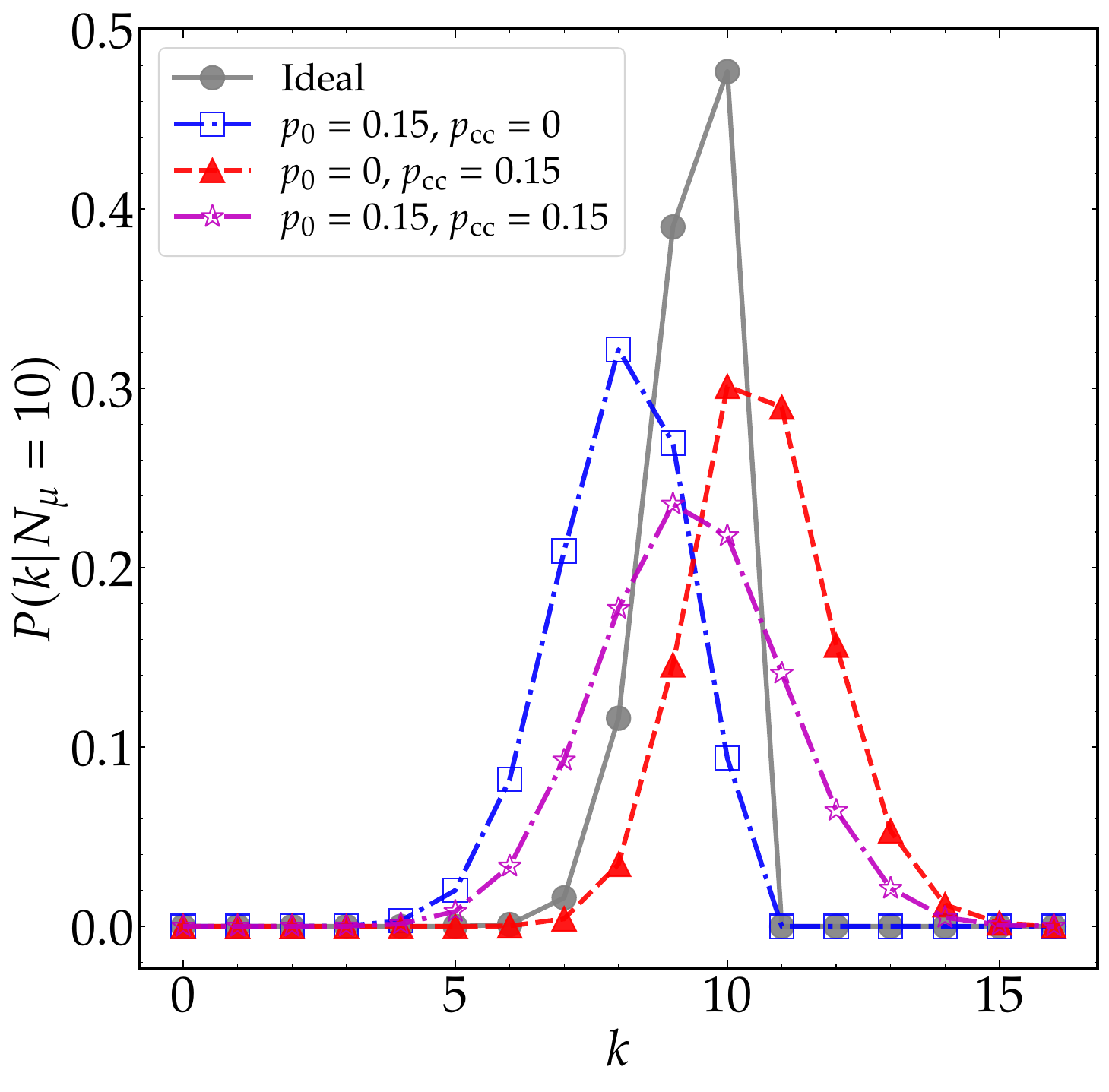}
    \caption{Probability distribution of $k$ for $\nmu=10$ (\eqnref{eq:prob_nmu}) for different values of $p_0$ and $p_\text{cc}$, setting $\lambda_1 = 0$.}
\label{fig:p_nmu_cases}
\end{figure}

In \figref{fig:p_nmu_cases}, we show $P(k \mid N_{\mu} {=} 10)$ for different values of $p_0$ and $p_\text{cc}$.
In the ideal case with $p_0 = p_\text{cc} = 0$ (circles), the distribution peaks around $k=\nmu$ and is constrained by $k \leq \nmu$, as previously shown in \figref{fig:p_nmu_ideal}.
This constraint remains evident when $(p_0 = 0.15, p_\text{cc} = 0)$ (empty squares), but the distribution now peaks at smaller values due to detector inefficiency.
When $(p_0 = 0, p_\text{cc} = 0.15)$ (triangles), the distribution is shifted toward larger values, and $k$ is no longer restricted to be smaller than $N_{\mu}$ as a consequence of corner-clipping.
Finally, when $p_0 = p_\text{cc} = 0.15$ (stars), the effects of detector inefficiency and corner-clipping partially compensate each other, resulting in a distribution that peaks around $\nmu$, as in the ideal case, but is significantly broader.


%

We now consider the impact of background signals.
Since this effect is independent of $N_{\mu}$, corner-clipping, and detector efficiency, we set $p_0 = p_\text{cc} = 0$ in the following for illustrative purposes.
This allows us to isolate the impact of background signals, controlled by $\lambda_1$, on the shape of the distribution.
Importantly, the qualitative behavior induced by $\lambda_1$ discussed below remains valid in the general case in which $p_0$ and/or $p_\text{cc}$ are non-zero.

\begin{figure}
    \centering
    \def\w{0.48}
\includegraphics[width=\w\textwidth]{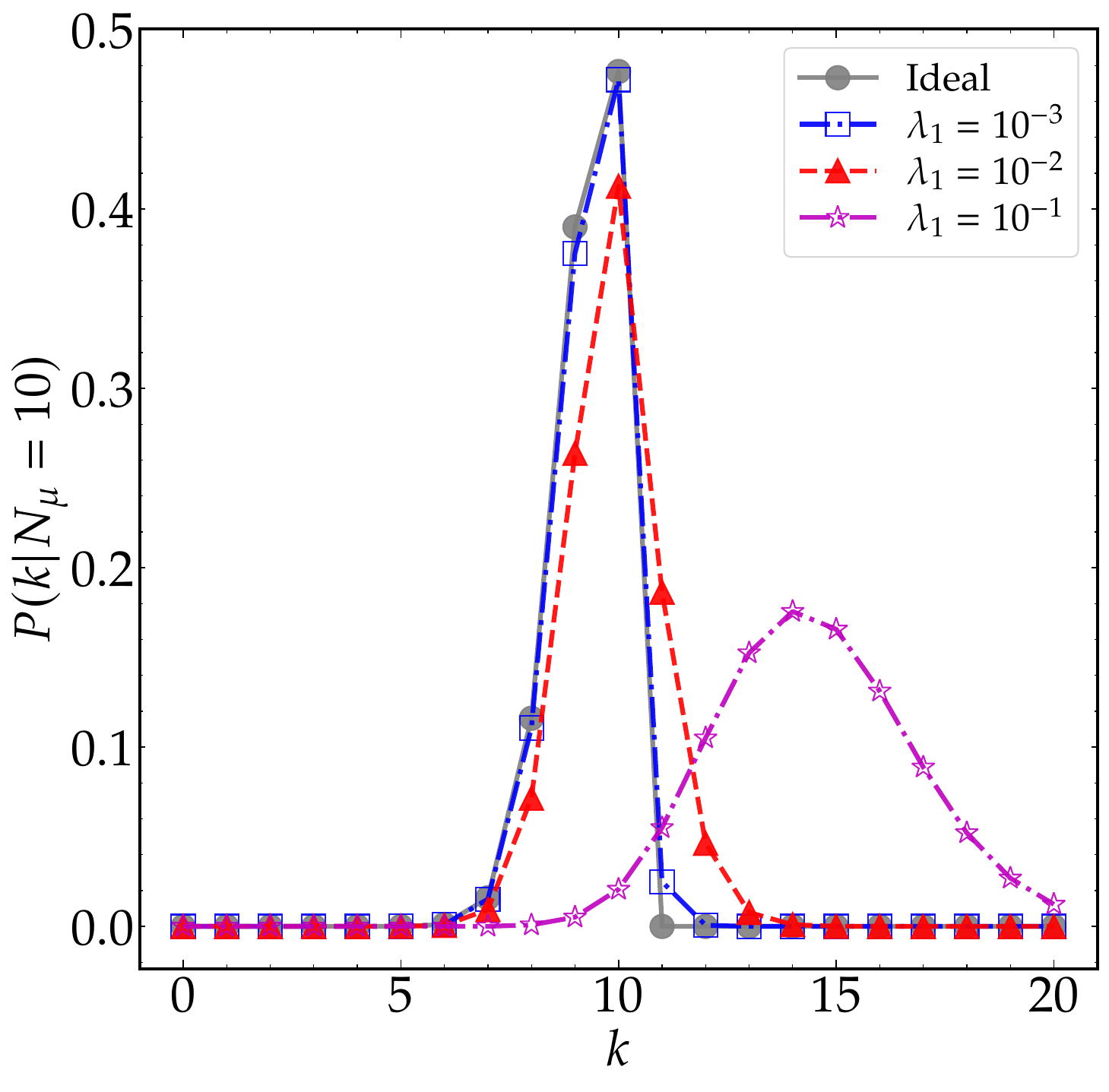}
    \caption{Probability distribution of $k$ for $N_{\mu}=10$ (\eqnref{eq:prob_nmu}) for different values of $\lambda_1$, setting $p_0 = p_\text{cc}= 0$. }
\label{fig:p_nmu_cases_bkg}
\end{figure}

In \figref{fig:p_nmu_cases_bkg}, $P(k \mid \nmu {=} 10)$ is shown for several values of $\lambda_1$. 
As $\lambda_1$ increases, background signals become more likely and the distribution progressively shifts toward larger values of $k$.
This behavior is generic and remains valid for arbitrary values of $p_0$ and $p_\text{cc}$.
The location of the peak of the distribution thus results from the competition between detector inefficiency (encoded in $p_0$), which shifts the distribution to smaller values of $k$, and corner-clipping ($p_\text{cc}$) and background ($\lambda_1$), which shift it toward larger values.

The inclusion of background in the model has important implications.
In the absence of background, the distribution for $\nmu = 0$ reduces to the trivial case $P(k=0 \mid \nmu {=}0) = 1$ and  $P(k\neq 0\mid \nmu {=}0) = 0$.
When background signals are included, this is no longer the case, since muons are no longer the only source of detector activity.
Evaluating \eqnref{eq:prob_nmu} for $\nmu = 0$ yields the non-trivial distribution (see \appref{app1})
\begin{equation}
    \label{eq:p_nmu_0}
    P(k \mid\nmu {=} 0) = 
    \binom{\ns}{k}\,
    p_\text{b}^{k}\,
    (1-p_\text{b})^{\ns-k},
\end{equation}
where $p_\text{b} = 1 - e^{-\lambda_1}$. 
This corresponds to a binomial distribution with $\ns$ trials and success probability $p_\text{b}$, which represents the probability that at least one background signal occurs in a given segment.
The existence of a non-trivial distribution $P(k \mid \nmu {=}0)$ implies an intrinsic detection limit, as discussed in \secref{sec:dist_mu}.

Taken together, the three detector effects play distinct roles: inefficiency shifts the distribution toward smaller values of $k$, whereas corner-clipping and background shift it toward larger values.
The resulting shape of the distribution is therefore governed by the interplay between these competing effects.

\subsection{Model assumptions and potential extensions}
\label{subsec:model_assumptions}
The extended model presented above provides a realistic and unified description of a segmented detector.
We now discuss the main assumptions underlying the model and their implications.

The first approximation concerns the spatial correlations introduced by corner-clipping muons.
In the analytical model, the two signals produced by a corner-clipping muon are treated as independent and uniformly distributed among the $\ns$ segments.
This treatment allows both signals to be assigned to the same segment, which is unphysical.
In reality, a corner-clipping muon necessarily activates two distinct neighboring segments.
The model therefore neglects both the requirement that the activated segments be distinct and the spatial correlation imposed by their adjacency.
Accounting explicitly for these constraints in the distribution of $k$ would considerably complicate the calculation and would likely preclude a simple closed-form expression.
In \secref{sec:monte_carlo_validation}, we assess the impact of neglecting these spatial correlations using Monte Carlo simulations and show that it has a negligible effect on the distribution of $k$ and on the performance of the resulting maximum-likelihood estimators and confidence intervals.



The second aspect concerns the interpretation of $p_\text{cc}$ in the presence of detector borders.
In a segmented detector, inner segments have two neighboring segments, whereas edge segments have only one.
Consequently, assuming that corner clipping toward either neighboring segment is equally probable for an inner segment, the corner-clipping probability of an edge segment is half that of an inner segment.
The parameter $p_\text{cc}$ entering the model should therefore be interpreted as an effective probability averaged over all detector segments.
Denoting by $p_\text{cc}^\text{inner}$ the corner-clipping probability for an inner segment and by $f_\text{edge}$ the fraction of edge segments, the detector-averaged probability is
%
%
\begin{equation}
p_\text{cc}
=
\left(1-\frac{f_\text{edge}}{2}\right)
p_\text{cc}^\text{inner}.
\end{equation}
For the UMD, each module is divided into two mechanically independent panels of 32 strips.
Since each panel has two end strips, four of the $\ns=64$ segments are edge segments, so that $f_\text{edge}=4/\ns$ and
\begin{equation}
\label{eq:pcc_inner_umd}
p_\text{cc}
=
\left(1-\frac{2}{\ns}\right)
p_\text{cc}^\text{inner}.
\end{equation}
Thus, the leading effect of detector borders on the average corner-clipping
probability can be absorbed into an effective,
detector-averaged value of $p_\text{cc}$, without modifying the form of the probability distribution.
Since the relative contribution of edge segments decreases as $1/\ns$, the border correction becomes progressively smaller for detectors with larger numbers of segments. 
For the UMD, the corresponding correction factor is $(1-2/\ns \simeq 0.97)$, indicating that border effects modify the effective corner-clipping probability by only about 3$\%$.


The third assumption concerns the maximum number of segments that can be activated by a single muon.
The model assumes that each muon activates at most two segments.
For sufficiently inclined trajectories, however, a muon may traverse and activate three or more segments, depending also on its azimuthal direction relative to the segment orientation.
The applicability of the model is therefore restricted to the angular range in which such configurations are negligible.
For the UMD of the Pierre Auger Observatory, which is designed for showers with zenith angles $\theta < 45^\circ$, the two-segment approximation provides an adequate description~\cite{DeJesus2026}.

Finally, we note that the framework can be readily extended to include additional mechanisms that, similarly to corner-clipping, produce one extra activated segment. 
In a real detector, effects such as crosstalk in the photodetector may generate this type of response.
Let $p_{\text{other}}$ denote the probability for such a process and $n_{\text{other}}$ the corresponding number of muons contributing through that process in a given event.
Provided that this response category is mutually exclusive with the other per-muon outcomes, it can be incorporated by replacing \eqnref{eq:Neff} with
\begin{equation}
M =  N_{\mu} - n_0 + n_{\text{cc}} + n_{\text{other}} + m,
\end{equation}
and extending the multinomial probabilities in \eqnref{eq:multinomial} such that
\begin{equation}
p_0+p_1+p_{\text{cc}}+p_{\text{other}}=1.
\end{equation}
This extension retains the assumption that the additional signal is independently distributed among the detector segments.
Processes that can occur simultaneously or introduce explicit spatial correlations would require a more general response model.

\section{Probability distribution of $k$ for fixed $\mu$}
\label{sec:dist_mu}

Having derived $\pknmu$, we now consider the case in which $\nmu$ is not fixed but fluctuates according to a Poisson distribution with mean $\mu$.
In air-shower reconstruction, the LDF predicts the expected number of impinging muons at each detector, $\mu$, rather than a particular realization $\nmu$.
The distribution $\pkmu$, which accounts for both these Poisson sampling fluctuations and the detector response for fixed $\nmu$, is therefore the relevant likelihood for LDF reconstruction.

We begin by reviewing the ideal-detector distribution $P_\text{ideal}(k \mid \mu)$ and its corresponding maximum-likelihood estimator, which provide a baseline for the full model in the limit $p_0=p_\text{cc}=\lambda_1=0$.
We then derive $\pkmu$ by marginalizing $\pknmu$ over the Poisson-distributed number of impinging muons.
Next, we interpret $\pkmu$ as a likelihood for $\mu$ and examine how detector effects modify its shape, its maximum, and the dynamic range over which $\mu$ can be reliably reconstructed.
Finally, we introduce a computationally efficient approximation and compare its maximum-likelihood estimators and confidence intervals with those obtained from the exact distribution.

\subsection{Ideal segmented detector}
As in the fixed-$N_{\mu}$ case, we begin by considering an ideal segmented detector.
Here, $N_{\mu}$ is treated as a Poisson-distributed variable with mean $\mu$.

For an ideal detector, each segment is activated with probability
$1-e^{-\mu/\ns}$, and the number of activated segments therefore follows a binomial distribution~\cite{ravignani2015},
\begin{equation}
\label{eq:prob_mu_base}
P_\text{ideal}(k \mid \mu) = {\ns \choose k} e^{-\mu}(e^{\mu/\ns} - 1)^k.
\end{equation}
For $k<\ns$, the corresponding maximum-likelihood estimator is
\begin{equation}
\label{eq:maximum_like_mu_ideal}
\hat{\mu}_\text{ideal} = -\ns \ln\left(1 - \frac{k}{\ns}\right).
\end{equation}
For $k=\ns$, the likelihood increases monotonically with $\mu$, and no finite maximum-likelihood estimate exists.

\subsection{Realistic segmented detector}

The distribution for a realistic detector is obtained by marginalizing
$\pknmu$ in \eqnref{eq:prob_nmu} over the Poisson-distributed number of
impinging muons, as defined in \eqnref{eq:prob_mu_marginalization}.
Using the finite-sum representation of $\pknmu$, this marginalization can
be carried out analytically, yielding (see \appref{app:pkmu_derivation})
%
%
%
\begin{equation}
\label{eq:prob_mu_full}
\begin{split}
P(k \mid \mu) ={}&
{\ns \choose k}
\sum_{j=0}^{k} (-1)^{k-j} {k \choose j} \\
&\times
\exp\Bigg\{
-\left(1-\frac{j}{\ns}\right)
\Bigg[
\ns\lambda_1 \\
&\qquad\qquad
+ \mu\left(
1-p_0+p_\text{cc}\frac{j}{\ns}
\right)
\Bigg]
\Bigg\}.
\end{split}
\end{equation}
Equation~\eqref{eq:prob_mu_full} is exact within the assumptions discussed in \secref{subsec:model_assumptions}.


The general behavior of $\pkmu$ as a function of $p_0$, $p_\text{cc}$, and $\lambda_1$ is qualitatively similar to that of $\pknmu$. 
Detector inefficiency shifts the distribution toward smaller values of $k$, whereas corner-clipping and background signals shift it toward larger values.
Consequently, the position of the peak is determined by the interplay between these competing effects. 
The model correctly reproduces both limiting cases: the background-only distribution of \eqnref{eq:p_nmu_0} for $\mu=0$, and the ideal-detector distribution of \eqnref{eq:prob_mu_base} for $p_0=p_\text{cc}=\lambda_1=0$.

We now consider \eqnref{eq:prob_mu_full} from the perspective of statistical inference. 
For an observed value of $k$, the distribution $\pkmu$ becomes a likelihood function for the parameter $\mu$. 
This likelihood forms the basis for the reconstruction of the muon density at a detector and, ultimately, of the muon LDF.

Unlike the ideal case, there is no closed-form expression for $\muhat$, the maximum-likelihood estimator of $\mu$ obtained from \eqnref{eq:prob_mu_full}. 
Although $\muhat$ can be determined numerically, it is useful to derive an analytical approximation that can be evaluated directly.
This approximation also provides insight into the dynamic range of the detector and, in particular, into the emergence of a lower detection threshold in the presence of background.

A natural approximation can be obtained from the expectation value of $k$, which is given by (see \appref{app:mean_var_k_mu})

\begin{align}
\label{eq:mean_k_mu_full}
\langle k \rangle (\mu) = \ns \left( 1 -  e^{-a} \right),
\end{align}

where 

\begin{equation}
\label{eq:a}
a = \frac{\mu}{\ns}\left[ 1 - p_0 + p_\text{cc} \left(1 - \frac{1}{\ns}\right) \right] + \lambda_1.
\end{equation}

By inverting \eqnref{eq:mean_k_mu_full} and replacing the expectation value $\langle k \rangle$ with the observed value $k$, we obtain the approximate estimator

\begin{equation}
    \label{eq:mu_hat_approx}
    \mutilde = -\ns \, \frac{\ln\left(1 - \frac{k}{\ns}\right) + \lambda_1}{1 - p_0 + p_\text{cc}\left(1 - \frac{1}{\ns}\right)}.
\end{equation}

We note that \eqnref{eq:mu_hat_approx} reduces to \eqnref{eq:maximum_like_mu_ideal} when $p_0=p_\text{cc}=\lambda_1=0$.
When all segments are active ($k=\ns$), the detector is said to be \emph{saturated}. 
In this regime, the likelihood no longer provides a finite estimate of $\mu$, and both $\muhat$ and its approximation $\mutilde$ diverge. 
This behavior defines the upper end of the detector dynamic range.
At this stage, \eqnref{eq:mu_hat_approx} should be regarded as a moment-based approximation to the maximum-likelihood estimator. In \secref{sec:approximate_dist}, we show that it is also the exact maximum-likelihood estimator of a simple approximate distribution.

\begin{figure}
    \centering
    \def\w{0.48}
\includegraphics[width=\w\textwidth]{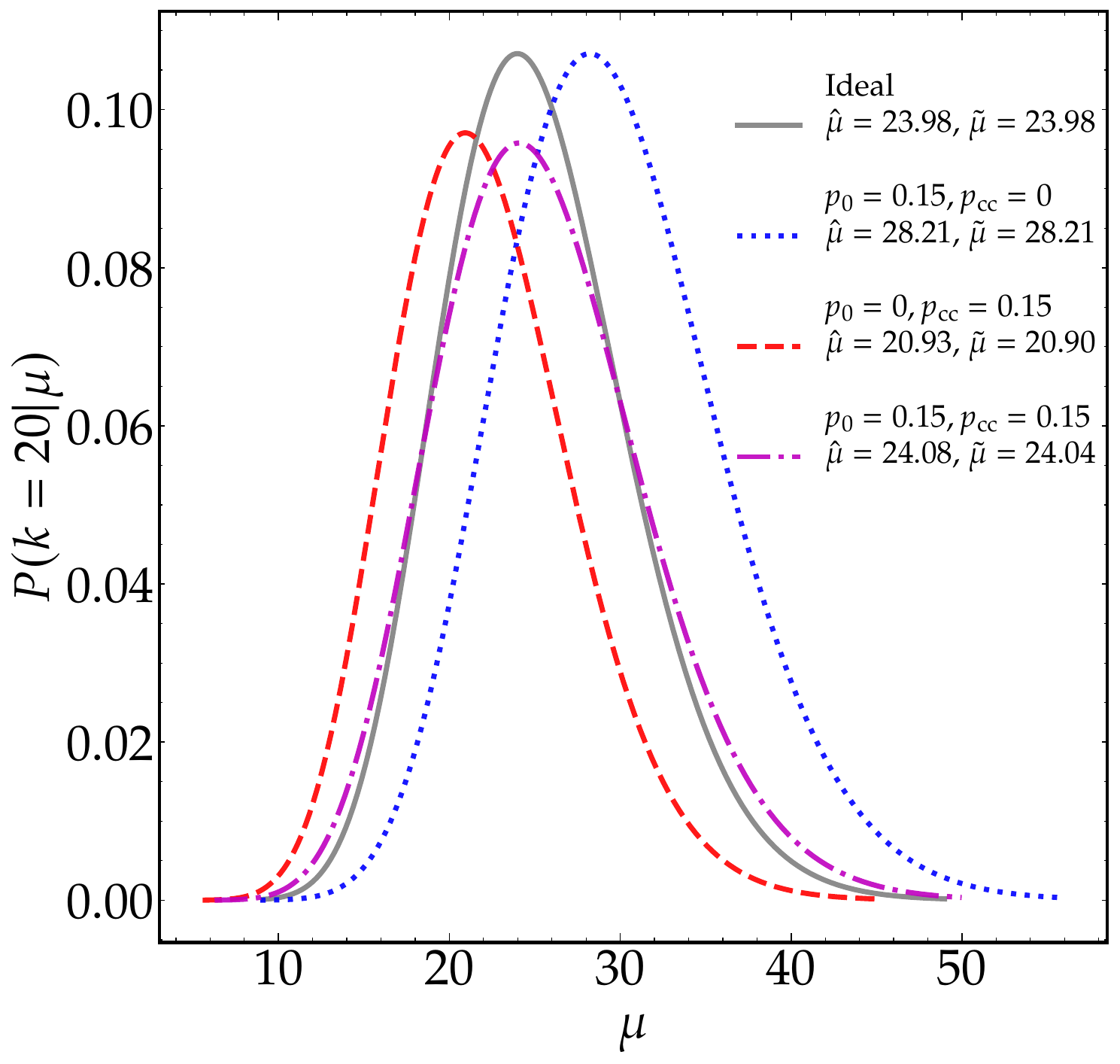}
    \caption{Likelihood as a function of $\mu$ for an observed value $k=20$ for different values of $p_0$ and $p_\text{cc}$, setting the background parameter $\lambda_1=0$. The maximum-likelihood estimator $\muhat$, obtained by maximizing \eqnref{eq:prob_mu_full}, and its approximation $\mutilde$ from  \eqnref{eq:mu_hat_approx}, are indicated for each case. }
\label{fig:like_scan_k20}
\end{figure}

We now examine the impact of $p_0$, $p_\text{cc}$, and $\lambda_1$ on the likelihood function. 
As in the discussion of the probability distributions, we first consider detector inefficiency and corner-clipping by setting $\lambda_1=0$, and then study the effect of background signals. 
In \figref{fig:like_scan_k20}, the likelihood defined by \eqnref{eq:prob_mu_full} is shown as a function of $\mu$ for an observed value $k=20$ and several combinations of $p_0$ and $p_\text{cc}$. 
Depending on the values of these parameters, different values of $\muhat$ are obtained for the same observed $k$.

%

When $p_0 > p_\text{cc}$ (dotted line), $\muhat$ is shifted toward larger values with respect to the ideal-detector case to compensate for muons that fail to activate detector segments. 
Conversely, when $p_0 < p_\text{cc}$ (dashed line), $\muhat$ is shifted toward smaller values, reflecting the additional segment activations induced by corner-clipping. 
Finally, when $p_0 = p_\text{cc}$ (dash-dotted line), the estimator remains close to the ideal-detector value, indicating that the effects of inefficiency and corner-clipping partially compensate each other, as discussed previously.
In all cases, the approximate estimator given by \eqnref{eq:mu_hat_approx} is also indicated in the figure, showing very good agreement with $\muhat$.

\begin{figure}
    \centering
    \def\w{0.48}
\includegraphics[width=\w\textwidth]{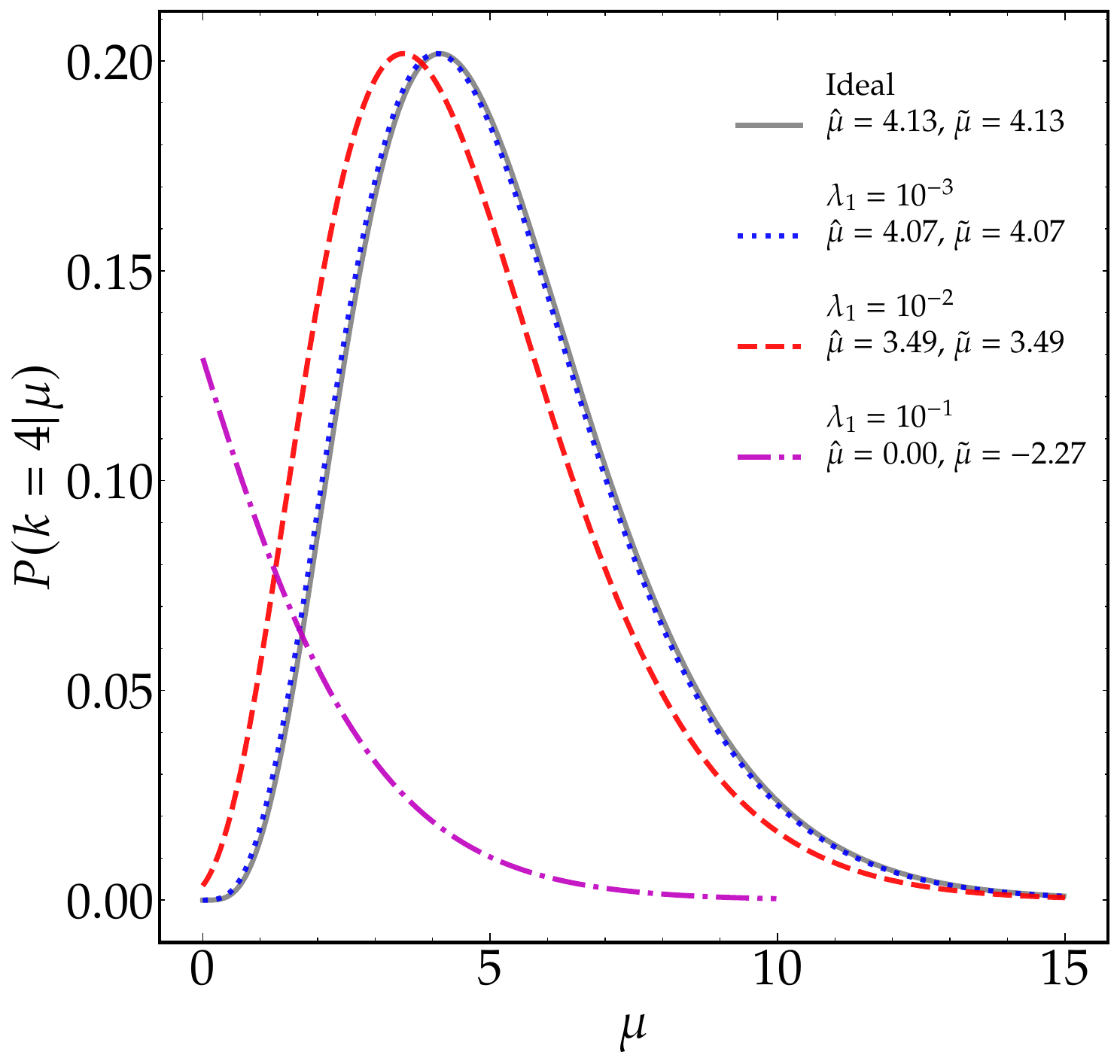}
    \caption{Likelihood as a function of $\mu$ for an observed value $k=4$ for different values of $\lambda_1$, setting $p_0=p_\text{cc}=0$. The maximum-likelihood estimator $\muhat$, obtained by maximizing \eqnref{eq:prob_mu_full}, and its approximation $\mutilde$ from  \eqnref{eq:mu_hat_approx}, are indicated for each case. }
\label{fig:like_scan_k4}
\end{figure}

We now turn to the background parameter.
In \figref{fig:like_scan_k4}, the likelihood is shown as a function of $\mu$ for an observed value $k=4$ and several values of $\lambda_1$, setting $p_0=p_\text{cc}=0$.
Different values of $\lambda_1$ lead to different likelihood shapes, with increasing $\lambda_1$ shifting the likelihood toward smaller values of $\mu$.
Notably, for $\lambda_1 = 10^{-1}$, the likelihood is maximized at $\muhat = 0$, even though the observed value of $k$ is non-zero.
In the same case, the approximate estimator $\mutilde$ yields a negative, non-physical value.
Both effects arise from the inclusion of background in the model, revealing that detector effects can fundamentally alter the range of $\mu$ values that can be reliably reconstructed.
This motivates the discussion of the detector dynamic range presented below.

%

\subsection{Implications for the dynamic range}
\label{sec:dynamic_range}

For physical values of the parameters, the denominator in \eqnref{eq:mu_hat_approx} is positive. 
Therefore, because of the overall minus sign, $\mutilde$ is positive only if the numerator is negative, which requires $k > k_\text{min}$, with
\begin{equation}
\label{eq:kmin}
k_\text{min} = \ns \left( 1 - e^{-\lambda_1} \right).
\end{equation}
This quantity corresponds to the expected number of active segments from background alone, i.e., the mean of the background-only binomial distribution in \eqnref{eq:p_nmu_0}.
When the observed $k$ is smaller than the expected background contribution, the likelihood is maximized at the physical boundary $\muhat=0$, while the approximate estimator becomes negative. 
Such observations do not provide a positive point estimate of the muon signal, although they may still constrain an upper limit on $\mu$.
The existence of $k_\text{min}$ therefore reduces the lower end of the detector dynamic range, since small values of $\mu$ are more likely to produce observations with $k < k_\text{min}$, where the detector loses sensitivity to the muon signal.

In general, for fixed values of $p_0$ and $p_\text{cc}$, the sensitivity to small values of $\mu$ decreases as $\lambda_1$ increases. 
Higher background levels require a larger number of muons to produce $k > k_\text{min}$ and exceed background-induced fluctuations. 
For fixed $\lambda_1$, the sensitivity to small $\mu$ also decreases with increasing $p_0$, since more muons are required to activate a given number of segments.
Conversely, the sensitivity improves with increasing $p_\text{cc}$, as corner-clipping enhances the probability of reaching $k \geq k_\text{min}$.

Detector effects also influence the largest values of $\mu$ that can be reliably reconstructed. 
Increasing either $\lambda_1$ or $p_\text{cc}$ causes the detector to saturate at smaller values of $\mu$. For a fixed $\mu$, both effects increase the probability of observing a saturated detector ($k=\ns$), thereby reducing the sensitivity at the upper end of the dynamic range. 
In contrast, increasing $p_0$ delays saturation and extends the range of $\mu$ values that can be reconstructed. 
The detector dynamic range is therefore determined by the interplay between background, inefficiency, and corner-clipping.

\subsection{Approximate distribution}
\label{sec:approximate_dist}
The exact distribution in \eqnref{eq:prob_mu_full} incorporates both
detector effects and Poisson sampling fluctuations in the number of
impinging muons.
Although it admits a closed finite-sum expression, its direct evaluation involves an alternating sum whose individual terms can be much larger than the final result. 
Depending on $k$ and the model parameters, cancellations among terms of comparable magnitude can lead to catastrophic cancellation and a substantial loss of numerical precision in a straightforward floating-point implementation.
Such numerical issues can be mitigated by suitable rearrangements of the sum or by using increased numerical precision, at the cost of additional implementation complexity.
Moreover, repeated evaluation of the exact expression for many detectors
and likelihood iterations is less convenient than evaluating a standard
probability distribution.

The structure of the exact distribution also provides limited immediate
insight into its statistical and inferential properties.
In particular, its dependence on the detector parameters does not lead
to a closed-form maximum-likelihood estimator or to a simple
interpretation of the likelihood shape.
A standard distribution with a physically interpretable effective
activation probability would instead provide a more transparent
description of the detector response and facilitate the interpretation
and implementation of the likelihood.
We therefore seek a simpler approximation that retains the main
statistical properties of the exact distribution.
Hereafter, we refer to \eqnref{eq:prob_mu_full} as the exact distribution
and, when regarded as a function of $\mu$ for an observed value of $k$,
as the exact likelihood.

The mean in \eqnref{eq:mean_k_mu_full} has exactly the form of the mean
of a binomial distribution with $\ns$ trials and success probability
$1-e^{-a}$.
To assess whether a binomial distribution can also reproduce the width
of the exact distribution, we consider the variance of $k$, which is
given by (see \appref{app:mean_var_k_mu})
\begin{equation}
\begin{split}
\label{eq:var_k_mu_full}
\mathrm{Var}[k](\mu)
={}&
\ns e^{-a}\left(1-e^{-a}\right) \\
&+
\ns(\ns-1)\left(e^{-2b}-e^{-2a}\right),
\end{split}
\end{equation}

where
\begin{equation}
\label{eq:b}
b =
\frac{\mu}{\ns}
\left[
1-p_0+
p_\text{cc}\left(1-\frac{2}{\ns}\right)
\right]
+\lambda_1.
\end{equation}
The first term in \eqnref{eq:var_k_mu_full} is the variance of a
binomial distribution with success probability $1-e^{-a}$, whereas the
second term is a correction to it.
The difference between $a$ and $b$ is
\begin{equation}
a-b=\frac{\mu p_\text{cc}}{\ns^2},
\end{equation}
so the correction originates entirely from corner clipping and vanishes
when $p_\text{cc}=0$.
This correction is always smaller than the binomial contribution and remains subdominant over the parameter range explored in this work.

Motivated by these observations, we approximate the exact distribution
by a binomial distribution with $\ns$ trials and success probability
$1-e^{-a}$,
\begin{equation}
\label{eq:prob_mu_approx}
\tilde{P}(k \mid \mu)
=
{\ns\choose k}
\left(1-e^{-a}\right)^k
e^{-a(\ns-k)}.
\end{equation}
This approximation exactly reproduces the mean value of $k$ and retains
the dominant contribution to its variance.
Moreover, it becomes identical to the exact distribution when
$p_\text{cc}=0$, as demonstrated in
\appref{app:pkmu_pcc0}.
Its unconstrained maximum-likelihood estimator is $\mutilde$; when the
physical condition $\mu\geq0$ is imposed, the estimator is bounded at
$\mu=0$.

\begin{figure}
    \centering
    \def\w{0.48}
\includegraphics[width=\w\textwidth]{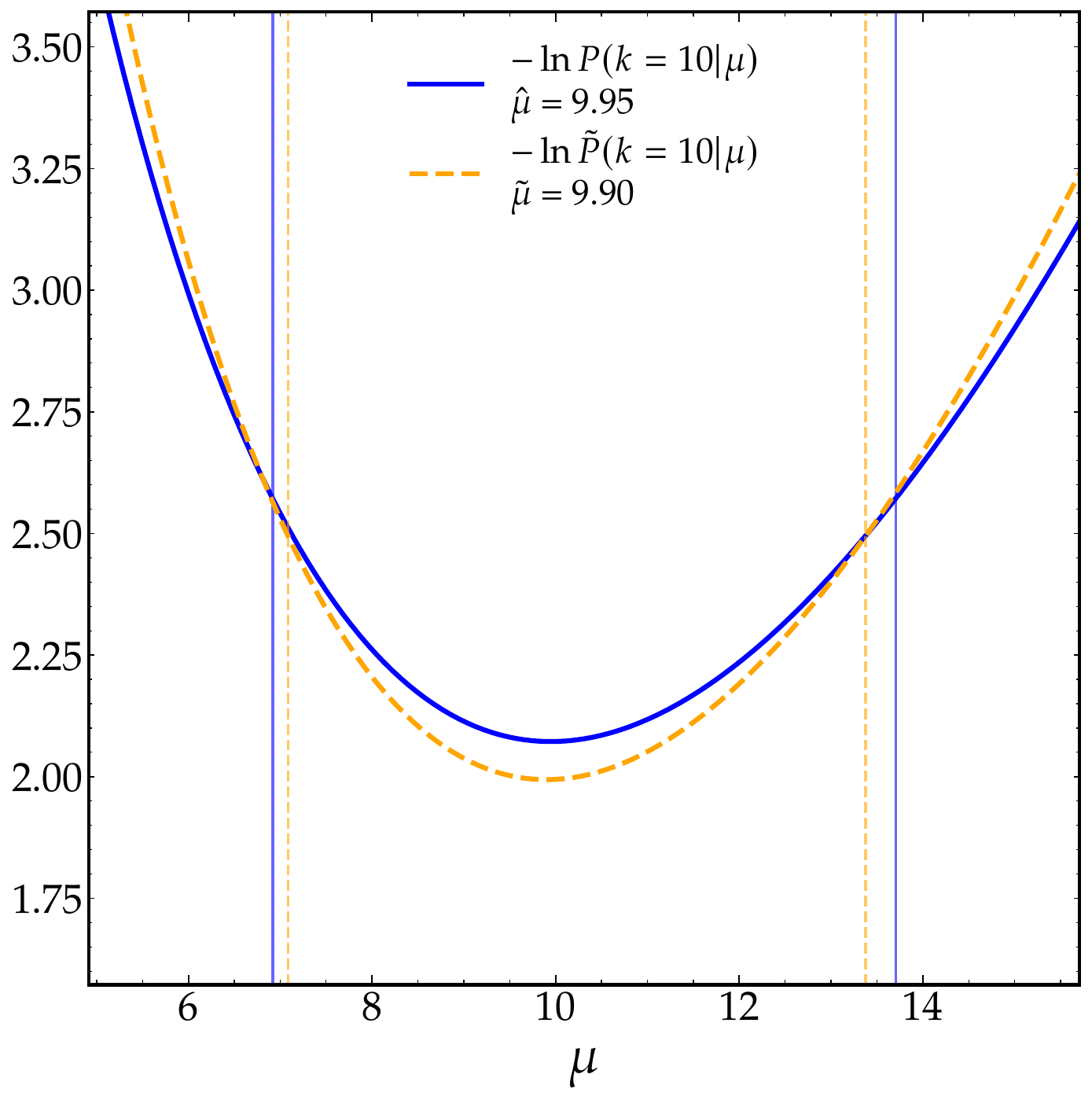}
    \caption{Negative log-likelihood as a function of $\mu$ for the exact (solid line) and approximate (dashed line) likelihoods, for an observed value of $k=10$. The vertical lines indicate the boundaries of the corresponding likelihood-ratio $1\sigma$ intervals. The parameters are set to $p_0=0$, $p_\text{cc}=0.1$, and $\lambda_1=0$.}
\label{fig:single_like_scan}
\end{figure}

We now assess how well the approximate likelihood reproduces the exact likelihood. 
\figref{fig:single_like_scan} illustrates the negative log-likelihoods obtained using the approximate and exact expressions as a function of $\mu$, for an observed value of $k=10$.
The parameters are set to $p_0 = 0$, $p_\text{cc} = 0.1$, and $\lambda_1 = 0$.
A non-zero value of $p_\text{cc}$ is deliberately chosen to highlight the differences between the two likelihoods, since $\tilde{P}(k \mid \mu)$ reduces to $\pkmu$ when $p_\text{cc} = 0$, as previously discussed.

The minima of the functions, corresponding to the maximum-likelihood estimators, are reported in the figure. 
Both estimators show excellent agreement, differing only by approximately 0.5$\%$.


The likelihood-based $1\sigma$ intervals $(\mu_1,\mu_2)$  are constructed using the standard likelihood-ratio prescription~\cite{Wilks1938}.
The interval boundaries correspond to the values of $\mu$ for which the negative log-likelihood increases by 0.5 relative to its minimum. 
For a regular one-parameter problem in the asymptotic regime, Wilks’ theorem associates this criterion with a nominal coverage of 68.3$\%$. 
In the present problem, however, the finite and discrete support of $k$, together with the physical boundary $\mu\geq0$, can lead to departures from the asymptotic result, particularly at small $\mu$. 
The actual frequentist coverage of these intervals is therefore evaluated explicitly in \secref{sec:monte_carlo_validation}.

The interval boundaries are indicated in the figure by solid and dashed lines for the exact and approximate likelihoods, respectively. 
The intervals obtained from the approximate likelihood are very similar to those of the exact likelihood, although they are systematically smaller, reflecting the fact that the approximate likelihood is slightly narrower than the exact one. 
This behavior is generally observed for $p_\text{cc} \neq 0$.

%
\begin{figure}
    \centering
    \def\w{0.48}
\includegraphics[width=\w\textwidth]{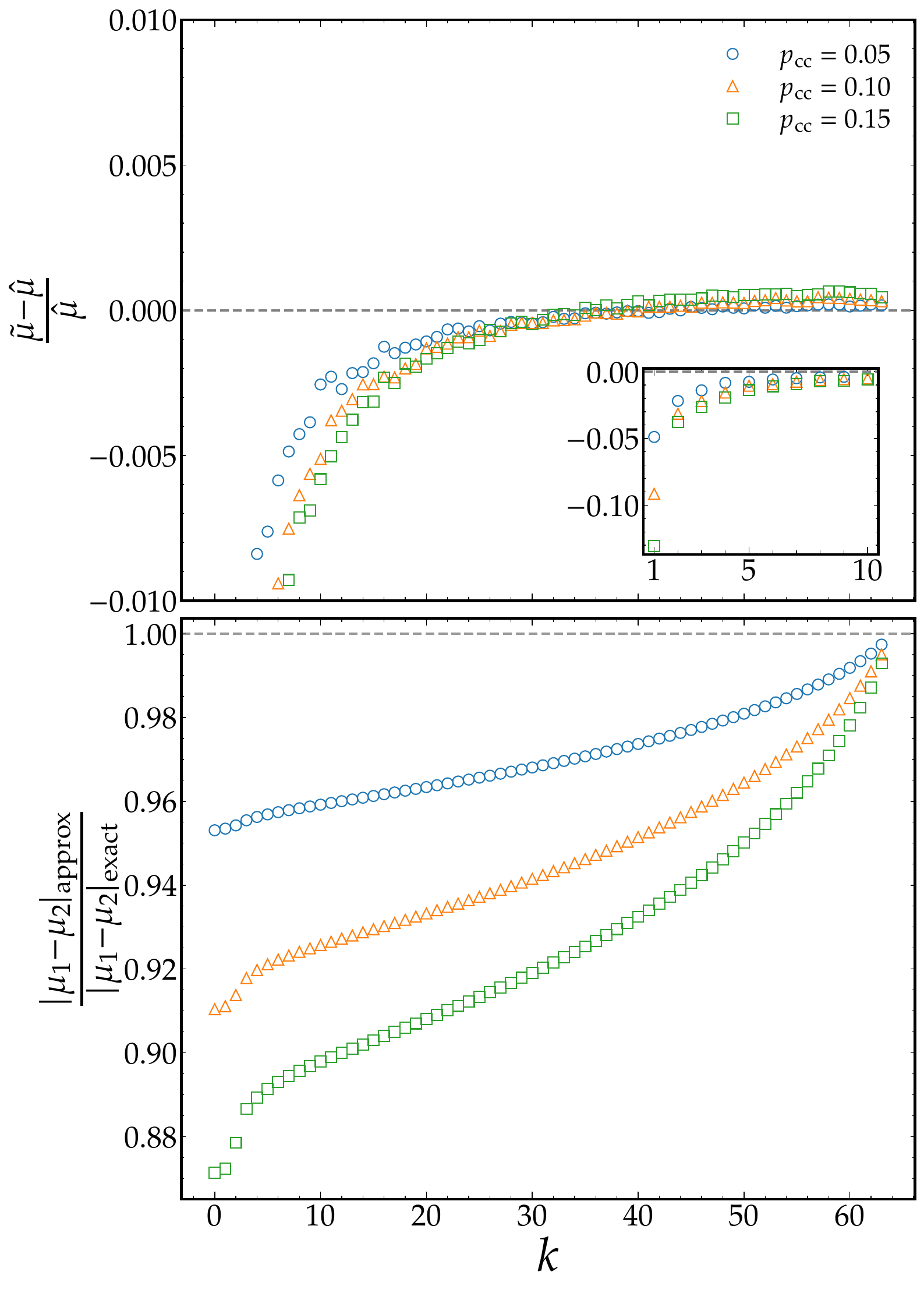}
    \caption{\emph{Top}: Relative difference between the approximate and exact maximum-likelihood estimators as a function of $k$ for the three non-zero values of $p_\text{cc}$ considered in this work. The inset panel shows a zoom into the region $1 \leq k \leq 10$. \emph{Bottom}: Ratio between the widths of the approximate and exact $1\sigma$ intervals as a function of $k$. The parameters are set to $p_0=0$ and $\lambda_1=0$.}
\label{fig:rel_res_two_pcc}
\end{figure}

In the top panel of \figref{fig:rel_res_two_pcc}, we show the relative difference between $\tilde{\mu}$ and $\muhat$ as a function of the observed occupancy $k$, for $p_\text{cc}\in \{0.05, 0.10, 0.15\}$. 
Again, we deliberately consider non-zero values of $p_\text{cc}$, since corner-clipping is the only detector effect that leads to differences between the exact and approximate likelihoods. 
In this example, we set $p_0=0$ and $\lambda_1=0$.

The agreement between $\mutilde$ and $\muhat$ improves as $k$ increases, reaching values within $1\%$ for $k > 4$ and $k > 7$ for $p_\text{cc}=0.05$ and $p_\text{cc}=0.15$, respectively. 
For small values of $k$, the absolute difference between the two estimators increases, reaching its maximum at $k=1$.

The discrepancy between the estimators becomes larger as $p_\text{cc}$ increases. 
For example, at $k=1$, the relative difference is ${\sim}5\%$ for $p_\text{cc}=0.05$, whereas it increases to ${\sim}13\%$ for $p_\text{cc}=0.15$, as shown in the inset of \figref{fig:rel_res_two_pcc}. 

The ratio between the lengths of the approximate and exact $1\sigma$ intervals as a function of $k$ is shown in the bottom panel of \figref{fig:rel_res_two_pcc}. 
For a fixed value of $p_\text{cc}$, the agreement between the exact and approximate intervals improves as $k$ increases. 
Nevertheless, the discrepancy becomes larger for increasing values of $p_\text{cc}$, similarly to the behavior observed for the estimator itself.
These behaviors reflect the fact that $\tilde{P}(k|\mu)$ progressively deviates from $P(k|\mu)$ as $p_\text{cc}$ increases.

The comparisons presented above quantify the differences between the approximate and exact likelihoods in terms of the resulting estimators and confidence intervals. 
However, these quantities alone do not determine the practical usefulness of the approximation. 
The relevant performance metrics are the bias of the estimator and the coverage of the confidence intervals. 
These quantities are evaluated in the next section using Monte Carlo simulations.

\section{Validation with Monte Carlo simulations}
\label{sec:monte_carlo_validation}

In this section, we use Monte Carlo simulations for two purposes.
First, we assess the impact of neglecting the spatial correlations introduced by corner-clipping muons, an approximation introduced in \secref{subsec:model_assumptions}. 
These correlations are explicitly included in the simulation but neglected in the analytical model, allowing us to quantify their impact on the distribution $\pkmu$.

Second, we evaluate the performance of the approximate likelihood $\tilde{P}(k \mid \mu)$. 
In particular, we compare the bias of the corresponding estimator and the coverage of its associated $1\sigma$ confidence intervals with those obtained using the full likelihood. 
The performance of the exact likelihood serves as the reference against which the approximate expression is evaluated.

\subsection{Simulation description}

For a fixed tuple $(\mu, p_0, p_\text{cc}, \lambda_1)$, the simulation proceeds as follows. 
Here, $p_\text{cc}$ denotes the effective corner-clipping probability averaged over all detector segments. 
To account explicitly for border effects, it is first converted into the corresponding probability for inner segments $p_\text{cc}^\text{inner}$ by inverting \eqnref{eq:pcc_inner_umd}.
This accounts for the fact that a corner-clipping candidate assigned to an edge segment produces a signal in a neighboring segment only for one of the two possible clipping directions.

The number of muons $\nmu$ is sampled from a Poisson distribution with mean $\mu$. 
The numbers of undetected muons ($n_0$), corner-clipping candidates ($n_\text{cc}^\text{cand}$), and ordinary single-segment muons ($n_1$) are then sampled from a multinomial distribution with probabilities $p_0$, $p_\text{cc}^\text{inner}$, and $1-p_0-p_\text{cc}^\text{inner}$, respectively.
The number of background signals, $m$, is sampled from the Poisson distribution given in \eqnref{eq:prob_noise}.
A random segment is then assigned to each of the $n_1 + n_\text{cc}^\text{cand}$ muons and to each background signal. 
For each corner-clipping candidate, an additional Monte Carlo decision is made according to whether the initially assigned segment is an inner or an edge segment. 
If it is an inner segment, one of its two neighboring segments is selected with equal probability and receives an additional signal.
If it is an edge segment, the two possible clipping directions are also selected with equal probability: one produces a signal in the only neighboring segment, whereas the other produces no additional signal. 
Thus, after averaging over all segment positions and clipping directions, the probability that an impinging muon activates two neighboring segments is $p_\text{cc}$. 
Finally, the number of segments containing at least one signal, $k$, is computed and stored as the simulation output.

The simulation is repeated $10^4$ times for each $(\mu,p_0,p_\text{cc},\lambda_1)$ configuration, yielding a histogram of $k$ with $10^4$ entries that provides an empirical estimate of $\pkmu$. 
The parameter $\mu$ is scanned over the range $0\leq\mu\leq319$ with a step size of $0.02$, while the detector-averaged corner-clipping probability and the remaining detector parameters take the values
\begin{align*}
p_0 &\in \{0,0.05,0.1,0.15\},\\
p_\text{cc} &\in \{0,0.05,0.1,0.15\},\\
\lambda_1 &\in \{0,10^{-3},10^{-2},10^{-1}\}.
\end{align*}

These parameter values were chosen to cover the range relevant for the UMD of the Pierre Auger Observatory, for which we expect $p_\text{cc}<0.12$~\cite{DeJesus2026}, $p_0\approx0$, and $\lambda_1\approx10^{-3}$~\cite{PierreAuger:2021calibration}, while also extending beyond the nominal operating conditions to test the estimators over a broad but still realistic parameter space.


\subsection{Validation of full distribution}

Differences between the analytical model and the simulations can arise only when $p_\text{cc} > 0$, because the simulation retains the spatial correlations  associated with corner-clipping signals that are neglected in \eqnref{eq:prob_mu_full}.
Consequently, the simulations follow the model exactly when $p_\text{cc}=0$. 
We therefore restrict the comparison to non-zero values of $p_\text{cc}$, for which differences between the model and the simulations are expected.

\begin{figure}
    \centering
    \def\w{0.48}
\includegraphics[width=\w\textwidth]{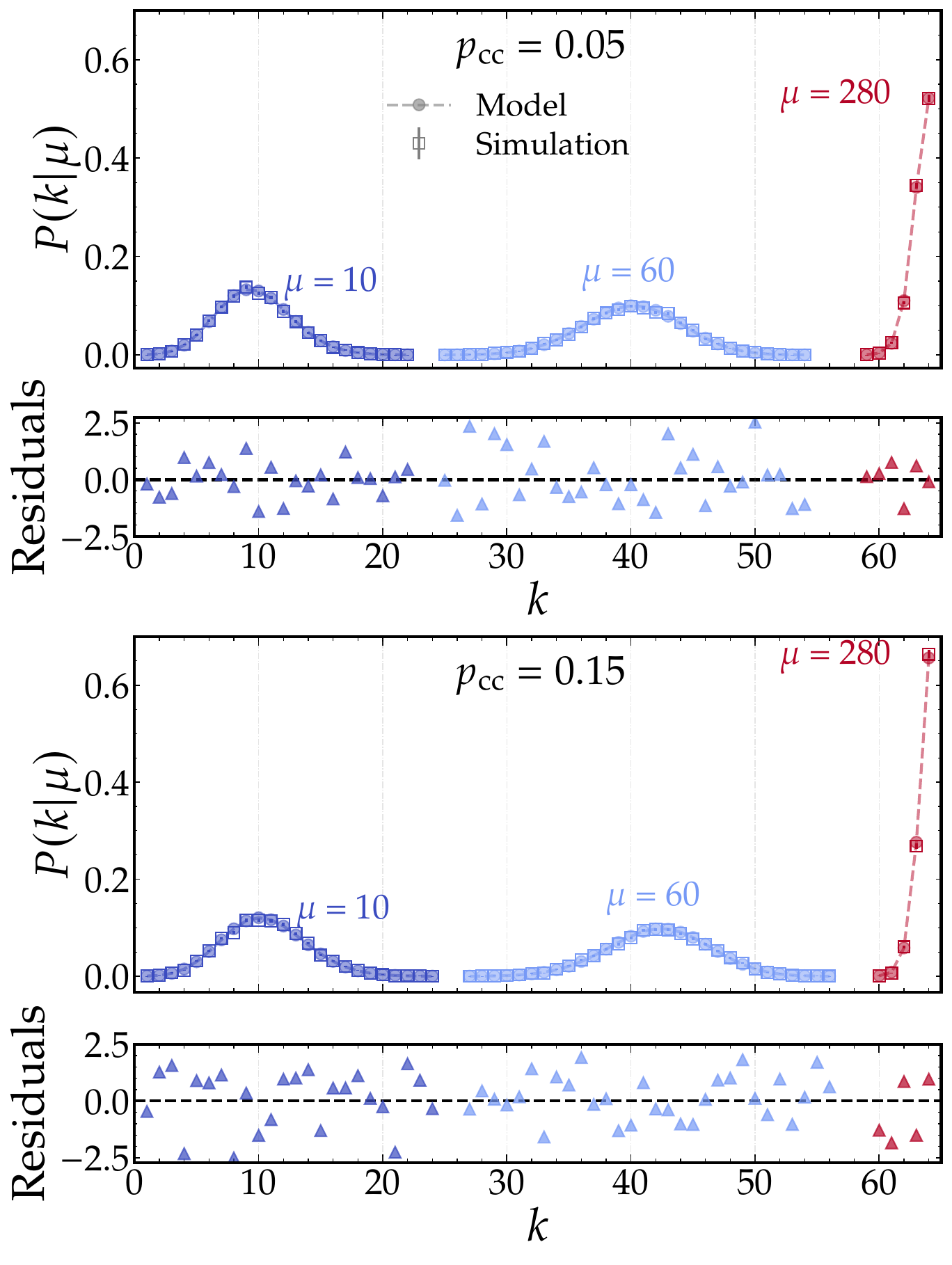}
    \caption{Comparison between the distribution of $k$ predicted by the model of \eqnref{eq:prob_mu_full} and that obtained from Monte Carlo simulations for representative values of $\mu$. For clarity, only values of $k$ for which the model predicts probabilities larger than $10^{-4}$ are shown. The lower plot in each panel shows the normalized residuals between the simulated and predicted distributions. The top panel corresponds to $p_\text{cc}=0.05$, whereas the bottom panel corresponds to $p_\text{cc}=0.15$. The remaining parameters are fixed to $p_0=0$ and $\lambda_1=10^{-3}$.}
\label{fig:residuals_dist}
\end{figure}

Figure~\ref{fig:residuals_dist} compares the distribution of $k$ predicted by the model of \eqnref{eq:prob_mu_full} with that obtained from Monte Carlo simulations for $p_\text{cc}=0.05$ (upper panel) and $p_\text{cc}=0.15$ (lower panel). 
In both cases, the model and the simulations are in excellent agreement. 
To quantify this agreement, the lower plot in each panel shows the normalized residuals, defined for each value of $k$ as
\begin{equation}
    \left(n_\text{sim}-n_\text{model}\right)/\sqrt{n_\text{model}},
\end{equation}
where $n_\text{sim}$ and $n_\text{model}$ denote the observed and expected number of counts in each bin, respectively. 
The residuals fluctuate around zero without an evident systematic structure and are broadly consistent with the statistical fluctuations of the simulations.
Similar agreement is observed for the remaining combinations of $p_0$, $p_\text{cc}$, and $\lambda_1$, indicating that the neglected correlations have no practically relevant impact over the range explored.

\subsection{Bias and coverage}
We now assess the performance of the approximate likelihood based on $\tilde{P}(k \mid \mu)$ relative to the exact likelihood based on $\pkmu$. 
We do so by comparing the bias of the corresponding maximum-likelihood estimator, $\mutilde$, with that of the exact estimator, $\muhat$, as well as the coverage of the confidence intervals derived from the two likelihoods.

\begin{figure}
    \centering
    \def\w{0.48}
\includegraphics[width=\w\textwidth]{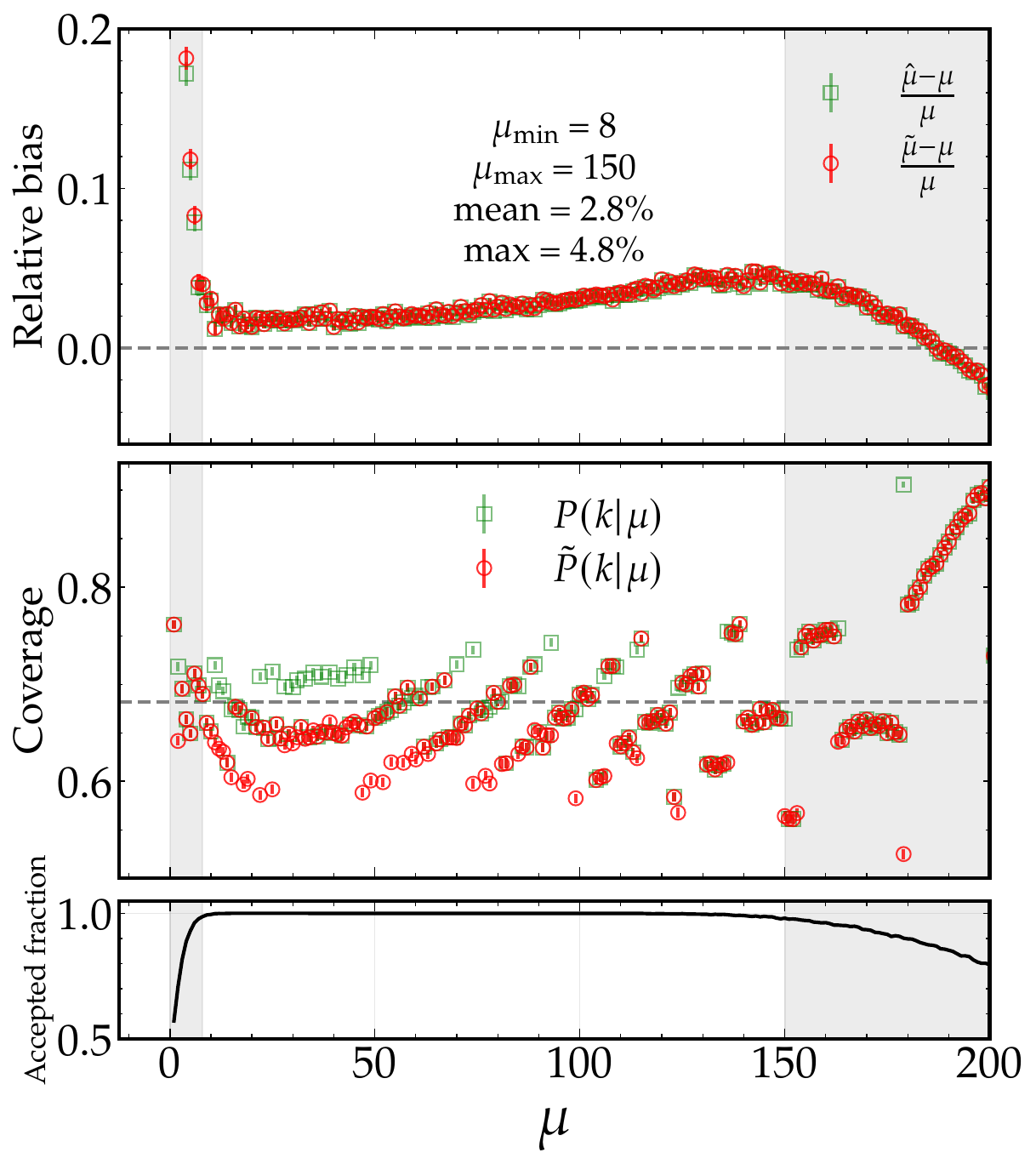}
    \caption{Assessment of the bias and coverage obtained with the exact (squares) and approximate (circles) likelihoods using toy Monte Carlo simulations for $p_0=0$, $p_\text{cc}=0.15$, and $\lambda_1 = 10^{-1}$. For clarity, only integer values of $\mu$ are shown. \emph{Top}: Relative bias of the exact and approximate maximum-likelihood estimators as a function of $\mu$. The minimum and maximum  values of $\mu$ that can be reliably reconstructed, $\mu_\text{min}$ and $\mu_\text{max}$, respectively, are indicated in the figure. The mean and maximum bias of $\mutilde$ within $\mu_\text{min} <\mu <\mu_\text{max}$ are also reported. \emph{Middle}: Coverage of the $1\sigma$ intervals obtained from the exact and approximate likelihoods as a function of $\mu$. The dashed horizontal line marks the nominal 0.683 value. \emph{Bottom}: Fraction of accepted toy Monte Carlo simulations as a function of $\mu$ ($k_\text{min} \leq k < \ns$). In all panels, the shaded area indicates the range where the accepted fraction drops below 0.98. }
\label{fig:bias_cover}
\end{figure}

In \figref{fig:bias_cover}, we compare the performance of the exact and approximate likelihoods in terms of bias and coverage for $p_0=0$, $p_\text{cc}=0.15$, and $\lambda_1=0.1$. 
These values were chosen to illustrate the most challenging region of the parameter space explored in this work. 
In particular, $p_\text{cc}=0.15$ corresponds to the largest corner-clipping probability considered, where the differences between the exact and approximate likelihoods are expected to be most pronounced.
Likewise, $\lambda_1=10^{-1}$ is the largest background parameter considered in this study, allowing us to highlight its impact on the minimum reconstructible value of $\mu$, as anticipated from the expression for $k_\text{min}$ discussed in \secref{sec:dynamic_range}.

The lower panel shows, as a function of $\mu$, the fraction of toy realizations  for which a finite, positive point estimate can be obtained, hereafter referred to as accepted realizations.
These are the realizations satisfying $k_\text{min}\leq k<\ns$. 
In all panels, the grey-shaded regions indicate the values of $\mu$ for which the fraction of accepted realizations falls below 0.98 and $\mu$ cannot therefore be reliably reconstructed. 
We adopt an accepted fraction of 0.98 as an operational criterion
for defining the reliable reconstruction range, with $\mu_\text{min}$ and $\mu_\text{max}$ (whose values are reported in \figref{fig:bias_cover}) denoting its lower and upper bounds, respectively.

The top panel shows the mean relative biases of $\mutilde$ and $\muhat$ as functions of $\mu$. 
The two estimators exhibit very similar biases over the full range considered, showing that $\mutilde$ closely reproduces the behavior of $\muhat$. 
Three regimes can be distinguished. 
For $\mu<\mu_\text{min}$, the bias changes rapidly because realizations with $k<k_\text{min}$ become increasingly likely as $\mu$ decreases. 
Since these realizations are excluded, the accepted sample becomes progressively dominated by upward fluctuations. 
Within the optimal operating range, $\mu_\text{min}<\mu<\mu_\text{max}$, the bias varies smoothly with $\mu$, reflecting the combined effects of the discrete detector response and the increasing relevance of finite segmentation. 
Finally, for $\mu>\mu_\text{max}$, saturation dominates and the estimators become biased toward lower values because only downward fluctuations satisfying $k<\ns$ remain reconstructible. 
The same qualitative behavior is observed for all parameter configurations explored.

The middle panel of \figref{fig:bias_cover} shows the coverage of the $1\sigma$ intervals obtained from the exact and approximate likelihoods as a function of $\mu$. 
Owing to the discrete nature of $k$, both intervals exhibit the irregular fluctuations typically associated with coverage studies for discrete observables. 
The approximate intervals yield coverages that are systematically equal to or lower than those of the exact intervals, consistent with their systematically smaller widths, as previously shown in \figref{fig:rel_res_two_pcc}. 
Nevertheless, the approximate likelihood reproduces the overall coverage behavior of the exact likelihood reasonably well.

\begin{figure}
    \centering
    \def\w{0.48}
\includegraphics[width=\w\textwidth]{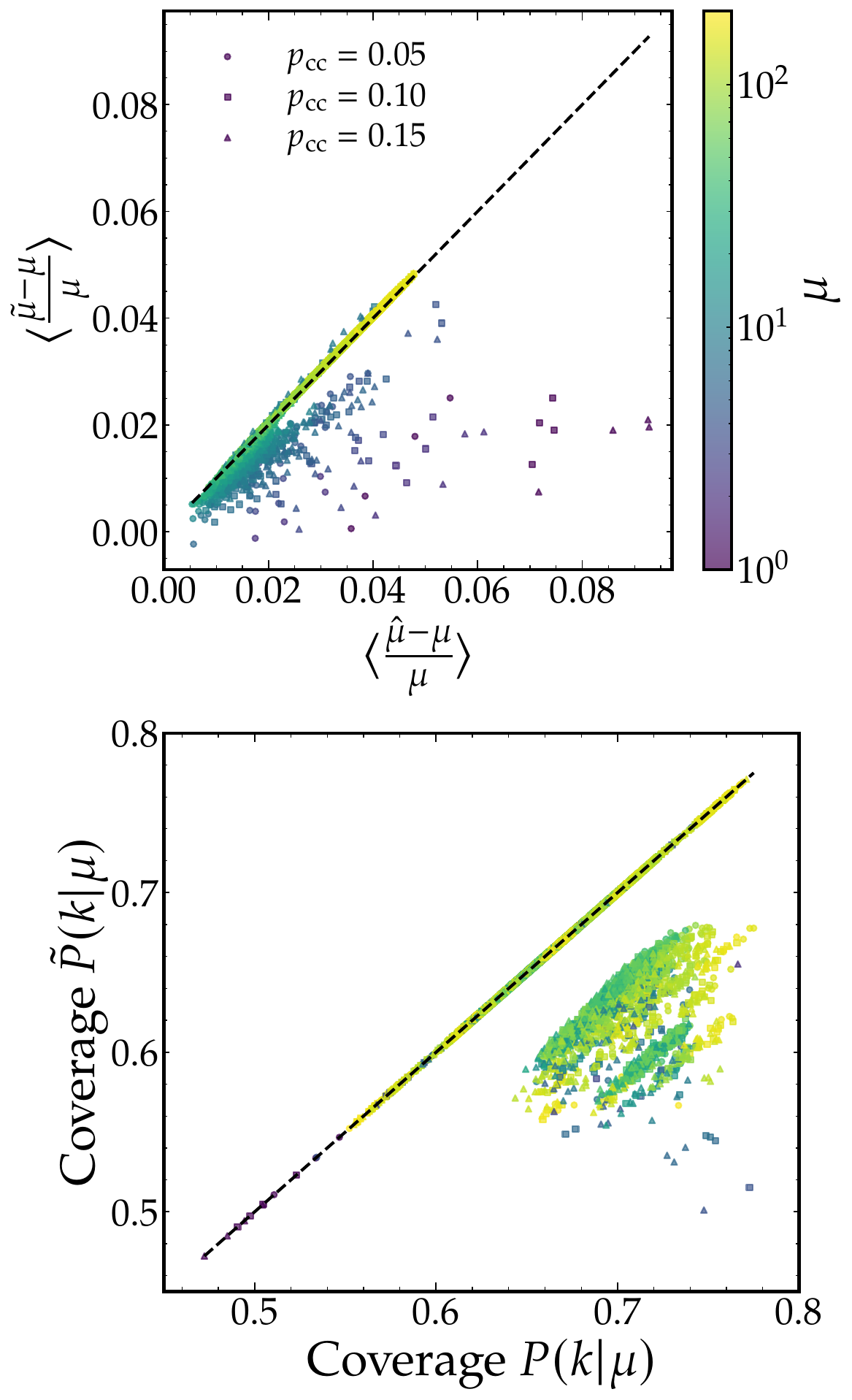}
    \caption{Comparison of the statistical performance of the approximate and exact likelihoods over the parameter space explored. Each point corresponds to one configuration $(\mu, p_0,p_\text{cc}, \lambda_1)$ within its reliable reconstruction range. Marker shapes indicate $p_\text{cc}$, while color encodes $\mu$. \emph{Top}: mean relative bias of the approximate estimator versus that of the exact estimator. \emph{Bottom}: coverage of the approximate likelihood-ratio interval versus that of the exact interval. The dashed lines indicate equality.}
\label{fig:bias_cover_all}
\end{figure}

The top panel of \figref{fig:bias_cover_all} compares the mean relative biases of $\mutilde$ and $\muhat$. 
Each point represents a parameter configuration $(\mu,p_0,p_\text{cc},\lambda_1)$. 
For each combination of $(p_0,p_\text{cc},\lambda_1)$, we include only values of $\mu$ within the reliable reconstruction range $\mu_\text{min}$ and $\mu_\text{max}$.
The value of $p_\text{cc}$ is encoded by the marker shape, while the color indicates $\mu$.

For most configurations, the biases of $\mutilde$ and $\muhat$ are very similar. Most points lie close to or below the identity line (dashed line), indicating that the bias of $\mutilde$ is generally equal to or lower than that of $\muhat$. 
This trend becomes more pronounced for large $p_\text{cc}$ and small $\mu$. 
In this regime, low values of $k$ are more likely, and $\mutilde$ is systematically smaller than $\muhat$, as shown in the top panel of \figref{fig:rel_res_two_pcc}.
Since $\muhat$ exhibits a positive bias in this region, the downward shift introduced by the approximate estimator partially compensates for this bias.

The lower panel of \figref{fig:bias_cover_all} compares the coverages of the $1\sigma$ intervals obtained from the approximate and exact likelihoods. 
The coverage of the approximate intervals never exceeds that of the exact intervals.
Most points lie close to the identity line, while several clusters appear below it. 
These deviations reflect the irregular fluctuations characteristic of coverage studies with discrete observables, as already discussed in connection with \figref{fig:bias_cover}.

Overall, the approximate likelihood closely reproduces the reconstruction performance of the exact likelihood, with very similar estimator biases and equal or lower confidence-interval coverages, while providing a numerically stable representation and a closed-form estimator.
The largest differences occur at small $\mu$ and large $p_\text{cc}$, where the approximate estimator is systematically lower than the exact one and partially compensates its positive bias.
The approximate confidence intervals are systematically narrower, which leads to coverages that are equal to or lower than those obtained with the exact likelihood.

These results indicate that the approximate likelihood is well suited for practical event-by-event reconstruction over the parameter range considered.
When maximum fidelity to the analytical detector-response model is required, however, the exact likelihood should be retained and evaluated with sufficient numerical precision to control cancellation errors in its finite-sum representation.

\section{Conclusions}
\label{sec:conclusions}

Segmented detectors are widely used to measure the muon content of extensive air showers.
For detectors with binary readout, the observable is the number of activated segments, $k$, and the reconstruction of the underlying muon density requires a probabilistic description of its relation to the number of muons reaching the detector.
Previous analytical models considered an ideal detector response and accounted only for signal pile-up.

In this work, we first derived the distribution $\pknmu$, which describes the detector response for a fixed number of impinging muons, $\nmu$.
This is the natural level at which to model detector effects, since detector inefficiency, corner-clipping muons, and background signals act on the actual particles and signals reaching the detector.
These effects were incorporated through the physically interpretable parameters $(p_0,p_\text{cc},\lambda_1)$, providing a unified and transparent description of the detector response.
For applications in which the actual number of impinging muons, $\nmu$, is the quantity of interest, $\pknmu$ provides the appropriate likelihood for statistical inference. 

For air-shower reconstruction, however, the relevant quantity is the expected number of impinging muons, $\mu$, predicted by the muon lateral distribution function.
We therefore derived $\pkmu$ by marginalizing $\pknmu$ over the Poisson fluctuations in $\nmu$.
The resulting distribution provides the likelihood contribution of an individual detector and constitutes the basic ingredient for reconstructing the muon LDF from an array of segmented detectors.

The distribution $\pkmu$ was then used to examine how detector effects influence the dynamic range, defined here as the range of $\mu$ over which the muon content can be reliably reconstructed.
At low $\mu$, background and detector inefficiency reduce the sensitivity to the muon signal, while corner clipping can partially compensate for these effects by increasing the observed segment multiplicity.
At high $\mu$, corner clipping and background accelerate detector saturation, whereas inefficiency delays saturation and extends the upper end of the reconstructible range.

Although the exact distribution admits a closed finite-sum expression, its direct evaluation involves an alternating sum that can suffer from catastrophic cancellation, particularly at large occupancies, and may require increased numerical precision.
Moreover, the corresponding likelihood does not yield a closed-form
maximum-likelihood estimator and its statistical structure is less
transparent than that of a standard distribution.
We therefore introduced a binomial approximation parameterized by an
effective segment-activation probability.
The approximation exactly reproduces the mean number of activated
segments and retains the dominant contribution to its variance, with the
remaining non-negative correction arising entirely from corner clipping
and always remaining smaller than the binomial contribution.
It also provides a closed-form estimator for $\mu$ and becomes identical
to the exact distribution when corner-clipping effects are absent.
For applications requiring the highest fidelity to the analytical
detector-response model, the exact likelihood should nevertheless be
used, with sufficient numerical precision to control cancellation errors.
The binomial approximation provides a numerically robust and readily
implementable alternative when the small deviations established in this
work are acceptable.

Dedicated Monte Carlo simulations were used for two complementary purposes.
First, they explicitly preserved the spatial correlations induced by corner-clipping muons and the detector-border geometry, allowing us to assess the impact of neglecting these features in the analytical model.
We found that their residual effect on $\pkmu$ is negligible over the parameter range considered.
Second, the simulations were used to compare the exact and approximate likelihoods.
The two approaches yield very similar reconstruction performance in terms of estimator bias and confidence-interval coverage, showing that the binomial approximation provides an accurate, numerically stable, and readily implementable representation of the exact model.

The framework developed here combines an exact probabilistic description of realistic detector effects with a small set of physically interpretable parameters.
At the same time, the binomial approximation provides a transparent and efficient likelihood for fitting muon LDFs in air-shower events recorded with segmented detectors.
Although the model was motivated by the Underground Muon Detector of the Pierre Auger Observatory, it is not limited to muon detection and applies to a broad class of segmented particle detectors with independent binary readout channels in which the particle content is inferred from the number of activated segments.
Detector-specific geometries or more complex intersegment correlations can be incorporated through suitable extensions of the response model.
By enabling a more realistic and statistically controlled reconstruction of muon densities, this framework can improve the use of segmented-detector data to constrain hadronic interaction models and thereby contribute to studies of the muon puzzle.

\begin{acknowledgement}
This work received support from the Comisión Nacional de Energía Atómica (CNEA), the Consejo Nacional de Investigaciones Científicas y Técnicas (CONICET), and the Universidad Nacional de San Martín (UNSAM) in Argentina, as well as from the Karlsruhe Institute of Technology (KIT) in Germany. Partial funding was provided by CONICET under grant PIP 2023-2025 GI 11220220100586CO.
\end{acknowledgement}

\section*{Data Availability}
The simulation datasets used in this study are available from the corresponding author upon reasonable request.

\section*{Code Availability}
The analysis code used in this work is available from the corresponding author upon reasonable request.

\appendix 
\section{Finite-sum expression for $\pknmu$}
\label{app:pknmu_simplified}

In this appendix, we show that the explicit expression in
\eqnref{eq:prob_nmu_brute} can be reduced to the finite-sum
form given in \eqnref{eq:prob_nmu}.

We use the following equivalent representations of the Stirling
numbers of the second kind,
\begin{equation}
\label{eq:stirling_alternative}
\begin{split}
k! \, S(M,k)
=&
\sum_{j=0}^{k}
(-1)^j {k\choose j}(k-j)^M
= \\
& \sum_{j=0}^{k}
(-1)^{k-j}{k\choose j}j^M.
\end{split}
\end{equation}
The equivalence of the two sums follows directly from the change
of summation index $j\rightarrow k-j$.

For a given configuration $(n_0,n_1,n_\text{cc},m)$, the total
number of signals distributed among the detector segments given by \eqnref{eq:Neff} can be alternatively written as
\begin{equation}
M=n_1+2n_\text{cc}+m.
\end{equation}
Using \eqnref{eq:stirling_alternative} in
\eqnref{eq:prob_nmu_brute}, and defining $z_j=\frac{j}{\ns}$, we obtain
\begin{equation}
\begin{split}
\label{eq:p_nmu_brute_aux}
\pknmu
&=
e^{-\Lambda}{\ns\choose k}
\sum_{j=0}^{k}
(-1)^{k-j}{k\choose j}
\\
&\!\!\! \!\!\!\!\!\!\!\times
\sum_{m=0}^{\infty}
\frac{(\Lambda z_j)^m}{m!}
\\
&\!\!\! \!\!\!\!\!\!\!\times
\sum_{n_0+n_1+n_\text{cc}=\nmu}
\frac{\nmu!}{n_0!n_1!n_\text{cc}!}
p_0^{n_0}
(p_1z_j)^{n_1}
(p_\text{cc}z_j^2)^{n_\text{cc}},
\end{split}
\end{equation}
where $\Lambda = \ns \lambda_1$.
Here, we have interchanged the order of summation. The sum over
$j$ and the multinomial sum are finite, while the series over
$m$ is absolutely convergent.

Using the exponential-series expansion, the sum over the number
of background signals becomes
\begin{equation}
\sum_{m=0}^{\infty}
\frac{(\Lambda z_j)^m}{m!}
=
e^{\Lambda z_j}.
\end{equation}
On the other hand, the multinomial theorem gives
\begin{equation}
\begin{split}
&\sum_{n_0+n_1+n_\text{cc}=\nmu} \!
\frac{\nmu!}{n_0!n_1!n_\text{cc}!}
p_0^{n_0}
(p_1z_j)^{n_1}
(p_\text{cc}z_j^2)^{n_\text{cc}}
\\
&\hspace{3cm}
=
\left(
p_0+p_1z_j+p_\text{cc}z_j^2
\right)^{\nmu}.
\end{split}
\end{equation}
Substituting these results into
\eqnref{eq:p_nmu_brute_aux}, we find
\begin{equation}
\begin{split}
\pknmu
&=
{\ns\choose k}
\sum_{j=0}^{k}
(-1)^{k-j}{k\choose j}
e^{-\Lambda(1-z_j)}
\\
&\quad\times
\left(
p_0+p_1z_j+p_\text{cc}z_j^2
\right)^{\nmu},
\end{split}
\end{equation}
which is the closed-form expression given in
\eqnref{eq:prob_nmu}.

\section{Finite-sum expression for $\pkmu$}
\label{app:pkmu_derivation}

In this appendix, we show how to obtain the expression of $\pkmu$ given in \eqnref{eq:prob_mu_full}. Inserting \eqnref{eq:prob_nmu} in \eqnref{eq:prob_mu_marginalization}, we obtain
\begin{equation}
\begin{split}
\label{eq:p_nmu_aux}
\pkmu
&= 
{\ns \choose k} e^{-\mu} \sum_{j=0}^{k} (-1)^{k-j} {k \choose j} e^{-\ns\lambda_1(1-z_j)}  \\ 
& \times \sum_{\nmu=0}^{\infty} \frac{\left[\mu(p_0 + p_1 z_j + p_\text{cc}z_j^2) \right]^{\nmu}}{\nmu!},
\end{split}
\end{equation}

where $z_j=j/\ns$. 
We have also interchanged the order of the
finite sum over $j$ and the absolutely convergent Poisson series
over $\nmu$. 
Using the series expansion of the exponential function,
\begin{equation}
    \sum_{\nmu=0}^{\infty} \frac{\left[\mu(p_0 + p_1 z_j + p_\text{cc}z_j^2) \right]^{\nmu}}{\nmu!} = e^{\mu(p_0 + p_1 z_j + p_\text{cc}z_j^2)}.
\end{equation}
Therefore,
\begin{align}
\pkmu
={}&
{\ns \choose k}
\sum_{j=0}^{k}
(-1)^{k-j}
{k \choose j}
e^{-\ns\lambda_1(1-z_j)}
\notag\\
&\times
\exp\left\{
-\mu\left[
1-p_0-p_1z_j-p_\text{cc}z_j^2
\right]
\right\}.
\end{align}
Using $p_1=1-p_0-p_\text{cc}$, the term in square brackets can
be factorized as
\begin{align}
1-p_0-p_1z_j-p_\text{cc}z_j^2
&=
(1-z_j)
\left(
1-p_0+p_\text{cc}z_j
\right).
\end{align}
It follows that
\begin{equation}
\begin{split}
\pkmu
={}&
{\ns \choose k}
\sum_{j=0}^{k}
(-1)^{k-j}
{k \choose j}
\\
&\times
e^{
-(1-z_j)
\left[
\ns\lambda_1
+
\mu\left(
1-p_0+p_\text{cc}z_j
\right)
\right]
},
\end{split}
\end{equation}

which, after substituting $z_j=j/\ns$, gives
\eqnref{eq:prob_mu_full}.

\section{Distribution of $k$ in the absence of muons}
\label{app1}

In this appendix, we show that \eqnref{eq:prob_nmu} reduces to
\eqnref{eq:p_nmu_0} when no muons impinge on the detector,
i.e., when $\nmu=0$.

Setting $\nmu=0$ in \eqnref{eq:prob_nmu}, the factor describing
the muon-induced signals becomes unity, and we obtain
\begin{equation}
\begin{split}
P(k \mid \nmu{=}0)
&=
{\ns\choose k}
\sum_{j=0}^{k}
(-1)^{k-j}{k\choose j}
e^{-\ns\lambda_1(1-j/\ns)}
\\
&=
{\ns\choose k}
e^{-\ns\lambda_1}
\sum_{j=0}^{k}
(-1)^{k-j}{k\choose j}
e^{\lambda_1 j}.
\end{split}
\end{equation}

Using the binomial identity
\begin{equation}
\label{eq:binomial_theorem}
\sum_{j=0}^{k}
(-1)^{k-j}{k\choose j}x^j
=
(x-1)^k,
\end{equation}
with $x=e^{\lambda_1}$, it follows that
\begin{align}
P(k \mid \nmu=0)
&=
{\ns\choose k}
e^{-\ns\lambda_1}
\left(e^{\lambda_1}-1\right)^k
\notag\\
&=
{\ns\choose k}
\left(1-e^{-\lambda_1}\right)^k
e^{-\lambda_1(\ns-k)}.
\end{align}
Defining $p_\text{b}=1-e^{-\lambda_1}$, we finally obtain
\begin{equation}
P(k \mid \nmu=0)
=
{\ns\choose k}
p_\text{b}^{\,k}
\left(1-p_\text{b}\right)^{\ns-k},
\end{equation}
which is the binomial distribution given in
\eqnref{eq:p_nmu_0}.










\section{Calculation of $\langle k \rangle$ and $\operatorname{Var}[k]$ for fixed $\nmu$}
\label{app:mean_var_k}

In this appendix, we derive the mean and variance of $k$ for a fixed number of impinging muons, $\nmu$.
These results are subsequently used to obtain the corresponding moments for fixed $\mu$.

Although the finite-sum expression in \eqnref{eq:prob_nmu} is more compact, the explicit representation in \eqnref{eq:prob_nmu_brute} is more convenient for this calculation.
In this form, the realistic detector response is expressed as a weighted average of ideal-detector distributions, allowing us to use directly the known expressions for the first and second moments of $k$ in an ideal segmented detector~\cite{supa2021}.

Using \eqnref{eq:prob_nmu_brute} and the definition of the mean, we obtain

\begin{equation}
\begin{split}
\label{eq:mean_k_definition}
\langle k \rangle &=  \sum_m  \sum_{n_0 + n_\text{cc}\leq N_{\mu}} P_\text{bg}(m) \, P(n_0, n_\text{cc} \mid N_{\mu}) 
\\ 
&
\times \sum_{k=0}^{\ns} k \, P_\text{ideal}(k \mid M).
\end{split}
\end{equation}
The innermost sum corresponds to the mean value of $k$ for an ideal segmented detector, which is given by~\cite{supa2021}
\begin{equation}
\label{eq:mean_k_base}
\langle k \rangle_\text{ideal}  = \ns \left[ 1 - \left(1-\frac{1}{\ns} \right)^{M}\right].
\end{equation}
After performing this substitution and some algebra, we obtain
\begin{equation}
\begin{split}
\label{eq:mean_k_definition2}
\langle k \rangle &=
\ns \Bigg[
1 -
\sum_m P_\text{bg}(m) \, \alpha^m \\
&\quad\times
\!\!\sum_{n_0 + n_\text{cc} \leq N_{\mu}}
P(n_0, n_\text{cc}\mid N_{\mu}) \, 
\alpha^{N_{\mu}-n_0+n_\text{cc}}
\Bigg],
\end{split}
\end{equation}
where $\alpha = 1 - 1 / n_\text{s}$.

Using the series expansion of the exponential function, it can be shown that $\sum_m P_\text{bg}(m)\alpha^m = e^{-\lambda_1}$.
%
%
On the other hand, using the multinomial theorem, it follows that
\begin{equation}
\begin{split}
\sum_{n_0 + n_\text{cc} \leq N_{\mu}}
P(n_0, n_\text{cc}\mid N_{\mu}) \,
\alpha^{N_{\mu}-n_0+n_\text{cc}}
&=
\\
\Big[
p_0 +
p_1\,\alpha
+
p_\text{cc}\alpha^2
\Big]^{N_{\mu}}.
\end{split}
\end{equation}

from which we finally obtain 

\begin{align}
\label{eq:mean_k_nmu_full}
\langle k \rangle = \ns \Big\{ 1 - e^{-\lambda_1}\Big[p_0 + p_1 \alpha + p_\text{cc}\alpha^2\Big]^{N_{\mu}}\Big\}.
\end{align}
For the calculation of the variance, we need $\langle k^2 \rangle$,
\begin{equation}
\begin{split}
\label{eq:k_squared_mean}
\langle k^2 \rangle 
&=
  \sum_m P_\text{bg}(m)
\sum_{n_0 + n_\text{cc}\leq N_{\mu}}
P(n_0, n_\text{cc} \mid N_{\mu}) 
\\
&\quad
\times \sum_{k=0}^{\ns} k^2 \, P_\text{ideal}(k \mid M).
\end{split}
\end{equation}
The innermost sum corresponds to the second moment of $k$ for an ideal detector which is given by~\cite{supa2021},
\begin{equation}
\begin{split}
\label{eq:k_squared_mean_ideal}
\langle k^2 \rangle_\text{ideal}
&=
\ns^2 - \ns(2\ns - 1)\alpha^M \, +
\\
&\quad
+\, \ns (\ns -1)\left(1-\frac{2}{\ns}\right)^M.
\end{split}
\end{equation}
Inserting \eqnref{eq:k_squared_mean_ideal} in \eqnref{eq:k_squared_mean} and using the series expansion of the exponential function and the multinomial theorem again, we obtain
\begin{align}
\label{eq:k_squared_mean2}
\langle k^2 \rangle =
  \ns \left(A_1 - A_2\right) + \ns^2\left(1 - 2A_1 + A_2 \right), 
\end{align}
with $A_i$  given by 
\begin{equation}
\label{eq:Ai_def}
A_i =
e^{-i\lambda_1}
\left[
p_0
+ p_1 \left(1 - \frac{i}{\ns}\right)
+ p_\text{cc}\left(1 - \frac{i}{\ns}\right)^2
\right]^{N_{\mu}} .
\end{equation}
Combining this result with the square of \eqnref{eq:mean_k_nmu_full}, we obtain the expression for the variance 
\begin{equation}
\label{eq:var_k_nmu_full}
\text{Var}[k] = \ns (A_1 - A_2) + \ns^2 (A_2 - A_1^2).
\end{equation}

\section{Calculation of $\langle k \rangle$ and $\operatorname{Var}[k]$ for fixed $\mu$}
\label{app:mean_var_k_mu}

In this appendix, we derive the mean and variance of $k$ for a fixed expected number of impinging muons, $\mu$.
As in the previous appendix, we use the explicit marginalization in
\eqnref{eq:prob_mu_marginalization} rather than the finite-sum expression
in \eqnref{eq:prob_mu_full}, since it allows us to obtain these moments
directly from the corresponding results for fixed $\nmu$ derived in
\appref{app:mean_var_k}.

The mean is given by
\begin{align}
\label{eq:mean_k_definition_mu}
\langle k \rangle
&=
\sum_{N_{\mu}=0}^{\infty}
P(N_{\mu} \mid \mu)
\sum_{k=0}^{\ns}
k \, P(k \mid N_{\mu}),
\end{align}
where $P(N_{\mu} \mid \mu)$ is a Poisson distribution with mean $\mu$ and $\pknmu$ is given by \eqnref{eq:prob_nmu}.

The innermost sum corresponds to the mean value of $k$ for fixed $\nmu$, given in \eqnref{eq:mean_k_nmu_full}. 
After performing this substitution and some algebra, we recover the expression given in \eqnref{eq:mean_k_mu_full}, using the series expansion of the exponential function.

To compute the variance, the second moment $\langle k^2\rangle$ is required:
\begin{equation}
\begin{split}
\label{eq:mean_k_squared_mu}
\langle k^2 \rangle
&=
\sum_{N_{\mu}=0}^{\infty}
P(N_{\mu} \mid \mu)
\sum_{k=0}^{\ns}
k^2 \, \pknmu.
\end{split}
\end{equation}
The innermost sum corresponds to the second moment of $k$ for fixed $\nmu$, given in \eqnref{eq:k_squared_mean2}.
After performing this substitution and using once again the series expansion of the exponential function, we obtain
\begin{align}
\label{eq:mean_k_squared_mu2}
\langle k^2 \rangle
&=
\ns \Big( e^{-a} - e^{-2b} \Big) + 
\ns^2 \Big(1 - 2e^{-a}  + e^{-2b}\Big),
\end{align}
where $a$ and $b$ are given by \eqnref{eq:a} and \eqnref{eq:b}, respectively.

Combining \eqnref{eq:mean_k_squared_mu2} with the square of \eqnref{eq:mean_k_mu_full}, we recover the expression for the variance given in \eqnref{eq:var_k_mu_full}.

\section{Distribution $\pkmu$ in the absence of corner clipping}
\label{app:pkmu_pcc0}

In this appendix, we show that the exact distribution in
\eqnref{eq:prob_mu_full} reduces to the binomial distribution
$\tilde{P}(k \mid \mu)$ when $p_\text{cc}=0$.

Setting $p_\text{cc}=0$ in \eqnref{eq:prob_mu_full}, we obtain
\begin{equation}
\begin{split}
\pkmu
={}&
{\ns\choose k}
\sum_{j=0}^{k}
(-1)^{k-j}{k\choose j}
\\
&\!\!\!\!\!\!\times
\exp\left[
-\left(1-\frac{j}{\ns}\right)
\left(
\mu(1-p_0)+\ns\lambda_1
\right)
\right].
\end{split}
\end{equation}
Defining $C=\mu(1-p_0)+\ns\lambda_1$, the distribution can be written as
\begin{align}
\pkmu
&=
{\ns\choose k}
e^{-C}
\sum_{j=0}^{k}
(-1)^{k-j}{k\choose j}
e^{Cj/\ns}.
\end{align}
Using the binomial theorem from \eqnref{eq:binomial_theorem} with $x=e^{C/\ns}$, we find
\begin{equation}
\begin{split}
\pkmu
&=
{\ns\choose k}
e^{-C}
\left(e^{C/\ns}-1\right)^k
\\
&=
{\ns\choose k}
\left(1-e^{-C/\ns}\right)^k
\left(e^{-C/\ns}\right)^{\ns-k}.
\end{split}
\end{equation}
Since $C / \ns = \mu(1-p_0) / \ns + \lambda_1$, defining
$q =1 - \exp\left[-\frac{\mu}{\ns}(1-p_0)-\lambda_1 \right]$, we finally obtain
\begin{equation}
\pkmu
=
{\ns\choose k}
q^k(1-q)^{\ns-k}.
\end{equation}
This is a binomial distribution with $\ns$ trials and success
probability $q=1-e^{-a}$, where $a$ is given by \eqnref{eq:a}
evaluated at $p_\text{cc}=0$.
It is therefore identical to the approximate distribution
$\tilde{P}(k \mid \mu)$ introduced in \secref{sec:approximate_dist}.

\bibliographystyle{JHEP}
\bibliography{references.bib}

\end{document}